\documentclass[aps,pra,twocolumn,superscriptaddress,amsmath,amssymb,amsfonts,10pt]{revtex4-2}

\usepackage{graphicx}
\usepackage{braket}
\usepackage{xcolor,color}
\usepackage[normalem]{ulem}
\usepackage[bookmarks=false,colorlinks=true,urlcolor=blue,citecolor=blue,linkcolor=blue]{hyperref}
\usepackage{bm}
\usepackage{microtype}
\usepackage{tikz}

\DeclareMathOperator{\mspan}{span}

\begin{document}

\title{Concomitant Entanglement and Control Criticality Driven by Collective Measurements}
\author{Thomas Iadecola}
\email{iadecola@iastate.edu}
\affiliation{Department of Physics and Astronomy, Iowa State University, Ames, IA 50011, USA}
\affiliation{Ames National Laboratory, Ames, IA 50011, USA}
\author{Justin H. Wilson}
\affiliation{Department of Physics and Astronomy, Louisiana State University, Baton Rouge, LA 70803, USA}
\affiliation{Center for Computation and Technology, Louisiana State University, Baton Rouge, LA 70803, USA}
\author{J. H. Pixley}
\affiliation{Department of Physics and Astronomy, Center for Materials Theory, Rutgers University, Piscataway, NJ 08854, USA}
\affiliation{
Center for Computational Quantum Physics, Flatiron Institute, 162 5th Avenue, New York, NY 10010, USA
}
\date{\today}

\begin{abstract}
Adaptive quantum circuits---where a quantum many-body state is controlled using measurements and conditional unitary operations---are a powerful paradigm for state preparation and quantum error correction tasks.
They can support two types of nonequilibrium quantum phase transitions: measurement-induced transitions between volume- and area-law-entangled steady states and control-induced transitions where the system falls into an absorbing state or, more generally, an orbit visiting several absorbing states.
Within this context, nonlocal conditional operations can alter the critical properties of the two transitions and the topology of the phase diagram.
Here, we consider the scenario where the measurements are \emph{nonlocal}, in order to engineer efficient control onto dynamical trajectories.
Motivated by Rydberg-atom arrays, we consider a locally constrained model with global sublattice magnetization measurements and local correction operations to steer the system's dynamics onto a many-body orbit with finite recurrence time.
The model has a well-defined classical limit, which we leverage to aid our analysis of the control transition.
As a function of the density of local correction operations, we find control and entanglement transitions with continuously varying critical exponents.
For sufficiently high densities of local correction operations, we find that both transitions acquire a dynamical critical exponent $z<1$, reminiscent of criticality in long-range power-law interacting systems.
At low correction densities, we find that the criticality reverts to a short-range nature with $z\gtrsim 1$.
In the long-range regime, the control and entanglement transitions are indistinguishable to within the resolution of our finite-size numerics, while in the short-range regime we find evidence that the transitions become distinct.
We conjecture that the effective long-range criticality mediated by collective measurements is essential in driving the two transitions together.
\end{abstract}

\maketitle

\section{Introduction}
Controlling chaotic dynamical systems is a longstanding problem relevant to a variety of real-world applications.
Control protocols for such systems entail measuring information about the system's state and, conditioned on this information, applying a feedback operation that pushes the system towards a target unstable fixed point or trajectory~\cite{ott1990controlling,antoniou_probabilistic_1997}.
It is natural to ask what is the minimal amount of intervention required to control a chaotic system.
Probabilistic control of chaos~\cite{antoniou_probabilistic_1996,antoniou_probabilistic_1997,antoniou_absolute_1998,antoniou_probabilistic_2000} provides a framework to answer this question by defining a hybrid discrete-time dynamics in which, at each time step, a control operation is applied stochastically with probability $p$ to overcome chaos.
In some archetypal examples of chaotic maps, control can emerge as a dynamical phase of matter upon crossing an inherently non-equilibrium phase transition at a critical control rate $p_c$.

With experimentally accessible noisy intermediate scale quantum (NISQ) devices, it is a timely and fundamental question to extend these ideas to monitored quantum systems.
In a prototypical class of dynamical maps, Refs~\cite{antoniou_probabilistic_1996,antoniou_absolute_1998} defined a probabilistic control protocol for the $\beta$-adic R\'enyi maps~\cite{renyi1957representations} which can also be used to control analogous quantum dynamics~\cite{Iadecola2023,LeMaire24,Allocca24,Pan24}.
The quantum control problem is enriched by measurements (necessary for the control protocol) which entail a backaction on the quantum state of the system, 
resulting in a phase transition in the late-time entanglement content of the quantum state. 
The entanglement transition can either precede or coincide with the (putatively) classical control transition depending on details of the entangling gates and control operation~\cite{SierantTurkeshi2023,LeMaire24,Pan24}.
Thus, probabilistic control of quantum systems fits into the broader context of adaptive quantum dynamics~\cite{Roy20,McGinley22,Sierant22a,Buchhold2022,ODea22,Ravindranath22,Friedman22b,MilekhinPopov2023,PiroliNahum2023,SierantTurkeshi2023,SierantTurkeshi2023a,Herasymenko23,Medina-Guerra23,Morales23,Langbehn23,Chertkov2023}, where measurement and feedback can drive critical phenomena separately witnessed by local order parameters and entanglement measures~\cite{li_quantum_2018,vasseur_entanglement_2019,li_measurement-driven_2019,nahum_measurement_2021,ippoliti_entanglement_2021,PotterVasseur2022a,FisherVijay2023}.
This class of adaptive quantum dynamics holds promise for applications in quantum technologies ranging from robust state preparation to quantum error correction.

The majority of monitored quantum circuits considered to date consist of short- to long-ranged unitary quantum gates and random local measurements.  
While measurements that project each local qubit fully destroy the entanglement of the many-body wavefunction, it is possible to perform a \emph{collective measurement} of a global property of the ensemble of spins (e.g., their total magnetization) which has an extensive number of possible measurement outcomes and does not fully disentangle the state. 
Long-range gates have a strong effect on the universality class of the measurement induced phase transition (MIPT)~\cite{Block22,Minato22,Muller22}, but it remains unclear if such a transition can persist in the presence of long-range or collective measurements and if so, what is its universal nature. 
Additionally, these collective measurements allows for a wider range of adaptive dynamics where we can feedback the results of these measurements to access a wider range fixed-point structures which could be used to overcome strongly entangled dynamics.

In this paper, we utilize adaptive quantum circuits as a framework to explore control and entanglement transitions in a setting that requires collective measurements; we explore this in a model of probabilistic quantum dynamics inspired by Rydberg-atom quantum simulators~\cite{Saffman10,Morgado21}.
The model consists of a chaotic circuit driven by kinetically constrained Rydberg-blockade dynamics~\cite{Iadecola20,Wilkinson20,Sellapillay22,Rozon22} that is stochastically interleaved with collective measurements and local control protocol (see Fig.~\ref{fig:full-circuit}).
The chaotic circuit (defined by $p=0$) possesses an unstable Hilbert-space orbit where the system cycles through a finite number of product states~\cite{Shiraishi18,Gopalakrishnan18,Iadecola20}.
The control protocol attempts to convert this unstable orbit into an attractor, i.e., a steady state that is ultimately reached regardless of the initial state, using collective measurements and local correction operations.
While our control protocol is applied stochastically, the dynamics within a protocol are necessarily adaptive.
We demonstrate the existence of control and entanglement transitions in this model with the aid of a classical limit in which the model maps onto a classical probabilistic cellular automaton (CA); this limit manifests a many-body generalization of classical probabilistic control.
With a variety of metrics, we estimate the locations and critical exponents of both transitions and find evidence that, in a large portion of the phase diagram, the transitions coincide and belong to the same universality class (although we cannot rule out the possibility that the transitions occur quite close to one another and with similar critical exponents).
Intriguingly, for a wide range of parameters, both transitions possess a dynamical critical exponent $z<1$, similar to MIPTs in monitored quantum dynamics with long-range unitary gates~\cite{Block22}.

The remainder of the paper is organized as follows.
In Sec.~\ref{sec: Circuit}, we define the probabilistic control model and its classical limit.
In Sec.~\ref{sec: Control Transition}, we study the control transition in the quantum model for a particular cut through the phase diagram and show where its critical properties match the transition in the classical model.
In Sec.~\ref{sec: Entanglement Transition}, we study the entanglement transition along the same cut through the lens of entanglement entropy, mutual information, and purification dynamics.
We consider the phase diagram in Sec.~\ref{sec: Phase}, and conclude in Sec.~\ref{sec: Conclusion} with a discussion and outlook.
Appendix~\ref{sec:Magnetization_Measurement} discusses how to implement the collective measurements underlying our control protocol in a qubit-only setup.
Appendices~\ref{sec: General Control} and \ref{sec: Sticky Orbits} discuss generalizations of our control protocol and the impact of finite-size effects, respectively.

\section{Model}
\label{sec: Model}

\begin{figure}[tb!]
    \centering
    \includegraphics[width=\columnwidth]{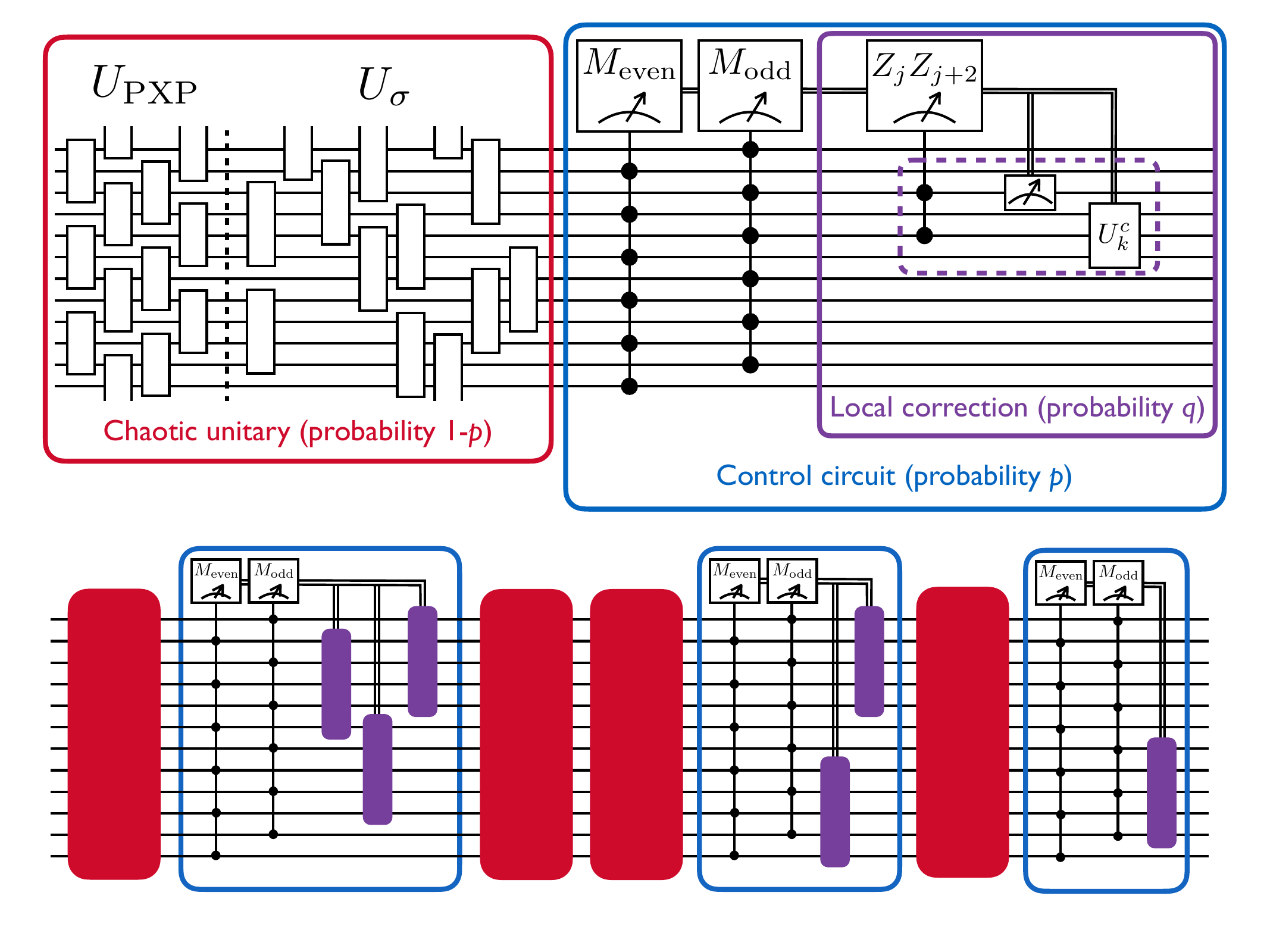}
    \caption{
    \textbf{Schematic depiction of the dynamics}. Dynamics defined in Sec.~\ref{sec: Model}.
    Top: At each time step, either a chaotic unitary circuit (red box, described in Sec.~\ref{sec: Circuit}) or a control circuit (blue box, described in Sec.~\ref{sec: Control}) is applied, depending on the outcome of a biased coin flip with bias $p$.
    As part of the control circuit, a local correction operation (purple box) is performed on each site $j$ with probability $q$.
    This operation acts within the five site region denoted by the purple box, since the correction operation $U^c_k$ acts on three sites and can be applied to site $k=j$ or $j+2$ depending on the various measurement outcomes.
    Bottom: An example of a random string of chaotic (red boxes) and control circuits. Within each blue box denoting a control circuit, the purple boxes represent the local correction operation applied with probability $q$ to each site.
    }
    \label{fig:full-circuit}
\end{figure}

The model we consider follows the high-level structure of probabilistic control of classical chaos~\cite{antoniou_probabilistic_1997}. 
This entails identifying an unstable orbit of the chaotic dynamics and stochastically applying a control operation that pushes the system onto this orbit. 
In particular, we study a one-dimensional system of $L$ qubits subject to a stochastic discrete-time dynamics in which, at each time step, with probability $1-p$ a ``chaotic" quantum circuit is applied, and with probability $p$ a control operation is applied.
The control operation is a hybrid quantum circuit involving measurements and local unitary feedback conditioned on the measurement outcomes.
We define the chaotic circuit in Sec.~\ref{sec: Circuit} and the control protocol in Sec.~\ref{sec: Control}; they are depicted schematically in Fig.~\ref{fig:full-circuit}. Importantly, each of them has a well-defined classical limit that we describe in Sec.~\ref{sec: Classical Limit}.

\subsection{Chaotic circuit}
\label{sec: Circuit}


The chaotic part of the dynamics is based on the ``PXP automaton" or ``Floquet PXP" circuit~\cite{Iadecola20,Wilkinson20}:
\begin{subequations}
\label{eq: UPXP-def}
\begin{align}
    U_{\rm PXP} = \prod_{j\text{ odd}} e^{-i\frac{\pi}{2}(PXP)_j} \prod_{j\text{ even}} e^{-i\frac{\pi}{2}(PXP)_j},
\end{align}
where
\begin{align}
\label{eq: PXP-def}
(PXP)_j = P^0_{j-1}X_jP^0_{j+1},\indent P^0_j = (1-Z_j)/2,
\end{align}
\end{subequations}
with $X_j,Z_j$ Pauli operators on site $j$ and corresponding local computational basis (CB) states $\ket{b_j}$ (with $b_j=0,1$) defined as eigenstates of $Z_j$ such that $Z_j \ket{b_j} = (-1)^{b_j+1}\ket{b_j}$. 
[We assume periodic boundary conditions (PBC) throughout this work.]
This circuit does not generate any entanglement when acting on CB states, and as such constitutes a reversible cellular automaton (CA); see Sec.~\ref{sec: Classical Limit} for more discussion on this type of dynamics.
The Hermitian generators in Eq.~\eqref{eq: PXP-def} perform a bit flip on site $j$ provided the bits on neighboring sites are in the $0$ state.
This constraint ensures that $U_{\rm PXP}$ never generates a bitstring containing a nearest-neighbor pair of 1s, provided the input state had no such pairs to begin with---this conservation law on the number of neighboring 1s is inspired by the Rydberg blockade in atomic physics~\cite{Jaksch00,Lukin01}.
Restricting to the zero-pair sector substantially reduces the dimension of the Hilbert space from $2^L$ to $\sim\varphi^L$, where $\varphi=(1+\sqrt{5})/2$ is the golden ratio.
For this reason, the zero-pair sector is sometimes referred to as the Fibonacci Hilbert space. 
In addition to its relevance to emerging quantum hardware, the reduced dimension of the Fibonacci Hilbert space is convenient for numerical simulations; we therefore focus on this subspace in the remainder of the paper.

The dynamics of the PXP automaton can be understood in terms of quasiparticles on top of the period-3 ``vacuum orbit": 
\begin{equation}
\label{eq: vacuum}
\begin{tikzpicture}[->,scale=1, baseline=(current  bounding  box.center)]
   \node (i) at (90:1cm)  {$\ket{0000\dots} $};
   \node (j) at (-30:1cm) {$ \ket{1010\dots}$};
   \node (k) at (210:1cm) {$\ket{0101\dots} $};

   \draw (50:1cm)  arc (50:-10:1cm);
   \draw (-50:1cm) arc (-50:-130:1cm);
   \draw (190:1cm) arc (190:130:1cm);
\end{tikzpicture}
\end{equation}
The quasiparticles are domain walls between any two states of the orbit, e.g.~$\dots00001010\dots$ or $\dots10100101\dots$, and they interact via time-delay scattering in such a way that the dynamics is fully integrable~\cite{Iadecola20,Wilkinson20}.
Any bitstring can be decomposed into a configuration of such quasiparticles, and each one lives on a unique closed orbit owing to the reversibility of the CA dynamics.
In the integrable model, typical orbits are of length $O(L)$, while the period-3 orbit \eqref{eq: vacuum} is the only orbit that appears consistently for any even system size.
States in the vacuum orbit are zero eigenvectors of the operator
\begin{align}
\label{eq:HZZ}
    H_{ZZ}=\frac{1}{L}\sum^L_{i=1}\frac{1-Z_{i}Z_{i+2}}{2},
\end{align}
which counts the total number of Ising domain walls on the even and odd sublattices of the chain.
The expectation value of $H_{ZZ}$ can thus be used as an order parameter for control onto this orbit.

\begin{figure}[tb!]
    \centering
    \includegraphics[width=\columnwidth]{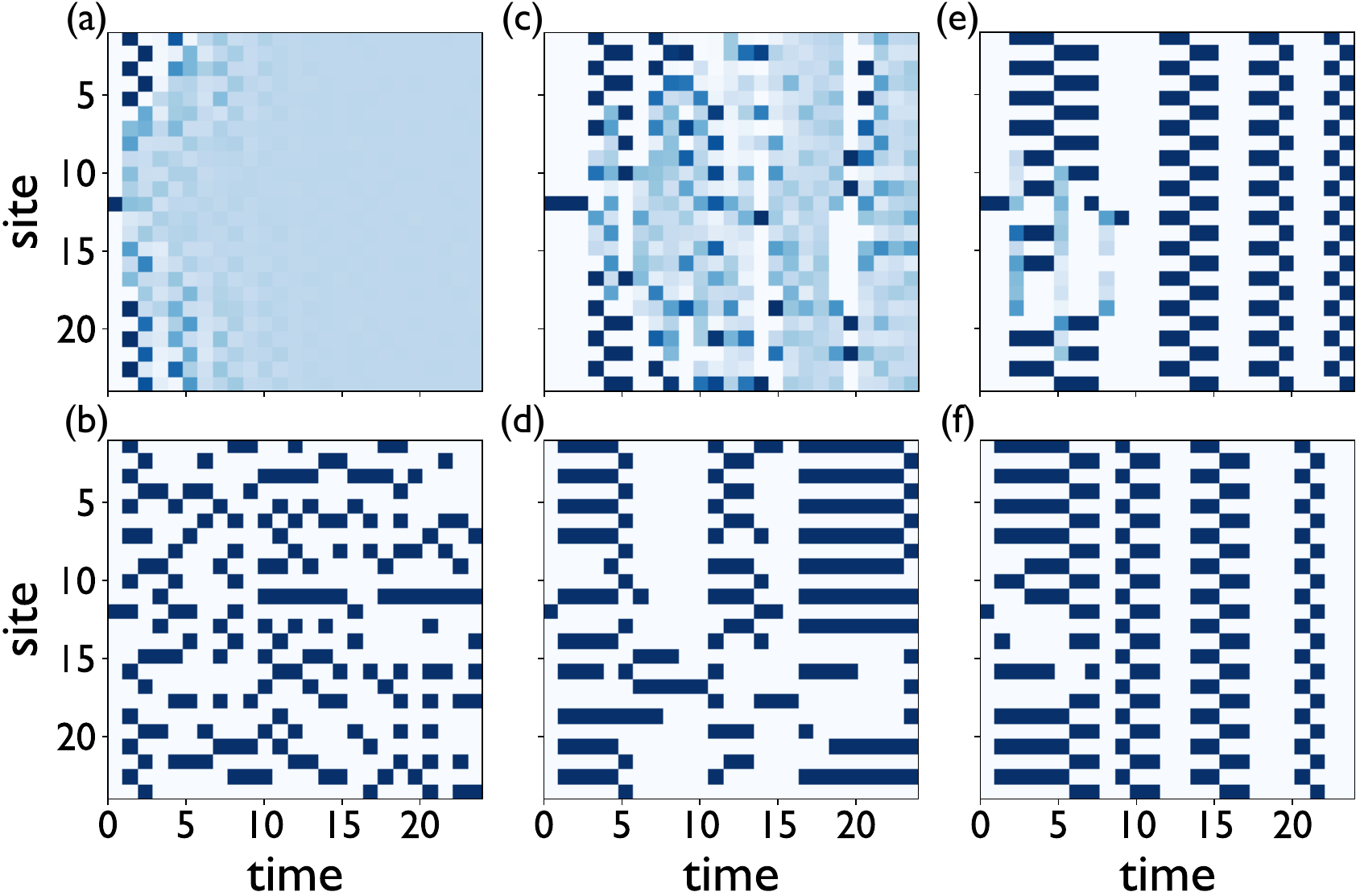}
    \caption{
    \textbf{Example spacetime dynamical trajectories}. For $p=0$ (a,b), $p=0.5$ (c,d) and $p=0.6$ (e,f) with $L=24$ and $q=0.2$.
    Each cell representing a spacetime point $(t,j)$ is colored according to the value of $\braket{Z_j}$ at time $t$ on a scale between $-1$ (white) and $+1$ (dark blue)
    All trajectories start from the same initial state, a CB state of the form $\ket{0\dots010\dots 0}$.
    Panels (a,c,e) use $\theta_Q=\pi/3$ in Eq.~\eqref{eq: U_sigma} and are representative of the quantum limit of the dynamics, with superpositions visualized by intermediate shades of blue.
    Measured regions that have locally collapsed to a CB state are clearly visible in panels (c) and (e).
    Panels (b,d,f) use $\theta_C=\pi/2$ and are representative of the classical limit of the dynamics.
    }
    \label{fig:example-evols}
\end{figure}

To restore a notion of chaotic dynamics, at each time step we follow the action of $U_{\rm PXP}$ by a random integrability-breaking circuit
\begin{align}
\label{eq: U_sigma}
    U_{\sigma}(\theta) = \prod_{j}e^{-i \frac{\theta}{2} P^0_{\sigma(j)-1}[X_{\sigma(j)}X_{\sigma(j)+1}+Y_{\sigma(j)}Y_{\sigma(j)+1}]P^0_{\sigma(j)+2}}.
\end{align}
Here, $\sigma$ is a permutation of the site indices $j=1,\dots,L$ drawn randomly at each time step, and $\sigma(j)$ is the image of site $j$ under this permutation.
Appending this circuit to $U_{\rm PXP}$ leads to chaotic dynamics from generic initial states.
However, since the generators of the local gates in Eq.~\eqref{eq: U_sigma} annihilate the states on the vacuum orbit \eqref{eq: vacuum}, the full circuit retains the vacuum orbit as a periodic trajectory for any value of $\theta$.
However, this trajectory is unstable in the sense that a single spin flip in the initial state takes the system off the orbit and leads to scrambling, as shown for a system of size $L=24$ with 
$\theta_Q=\pi/3$ and $\theta_C=\pi/2$
in Fig.~\ref{fig:example-evols}(a) and (b), respectively.

The red box in Fig.~\ref{fig:full-circuit} contains a schematic of one instance of the full chaotic circuit $U_{\sigma}(\theta)U_{\rm PXP}$. 
We emphasize that this circuit can be realized in Rydberg atom quantum simulators: 
\begin{itemize}
    \item $U_{\rm PXP}$ can be implemented by driving the even and odd sublattices of a 1D Rydberg chain in the nearest-neighbor blockade limit. 
    This can be achieved either with local Rabi driving, or with global driving in a two-species Rydberg array~\cite{Singh22,Singh23,Anand24}.
    \item The four-qubit gates entering $U_{\sigma}(\theta)$ can be implemented using local Rabi driving and local detuning of the Rydberg state~\cite{Koyluoglu24,MaskaraUnpublished}.
\end{itemize}

\subsection{Control Protocol}
\label{sec: Control}

\begin{figure}[tb!]
    \centering
    \includegraphics[width=0.8\columnwidth]{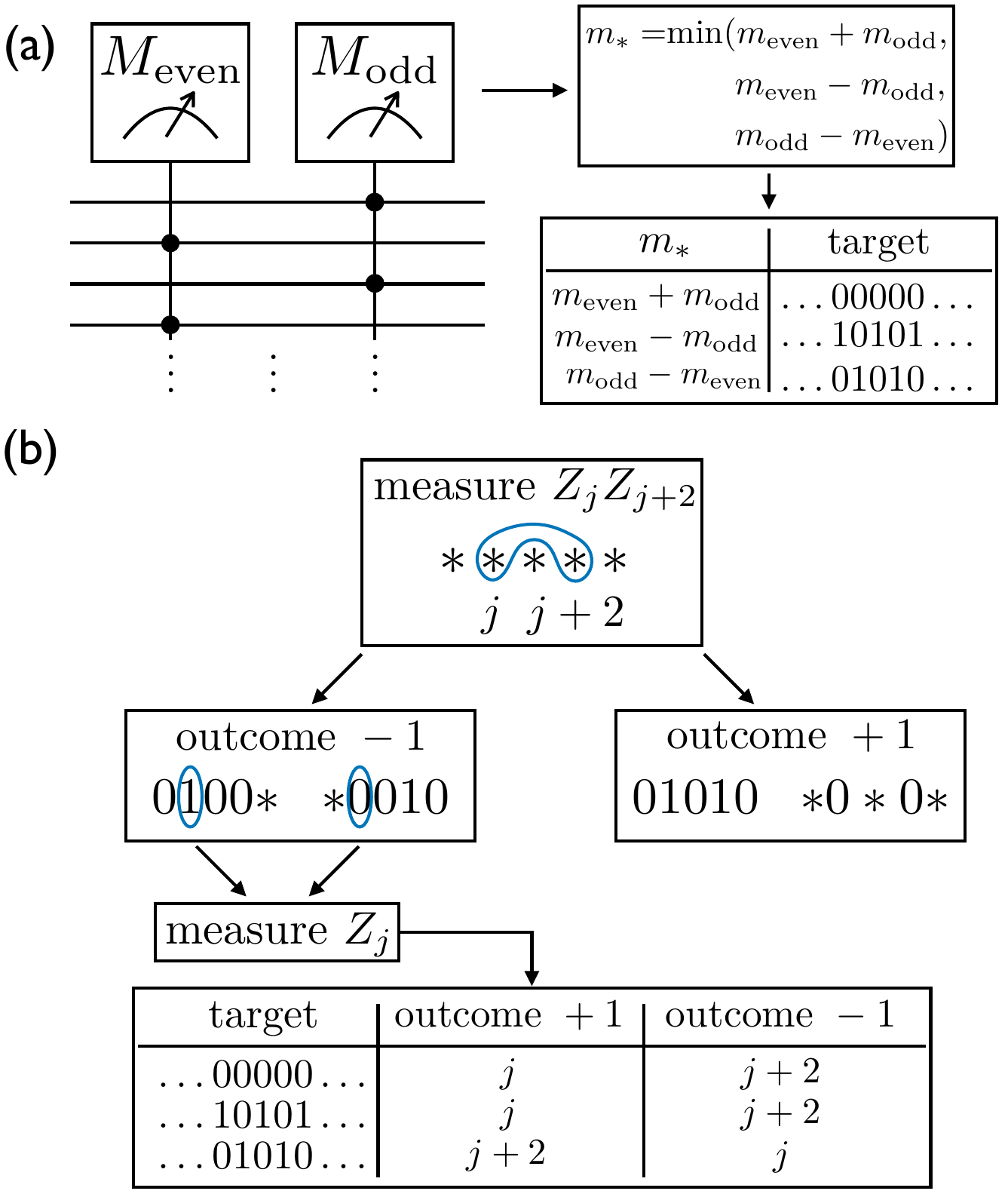}
    \caption{
    \textbf{Flowcharts for the two stages of the control protocol}. Protocol described in Sec.~\ref{sec: Control} and visualized in the blue box in Fig.~\ref{fig:full-circuit}.
    (a) Sublattice magnetization measurement outcomes $m_{\rm even(odd)}$ are used to calculate $m_*$, which determines the target state of the local correction step.
    (b) Flowchart for the local correction step (purple box in Fig.~\ref{fig:full-circuit}, applied with probability $q$ at each site $j$).
    First, a measurement is performed to see whether a sublattice domain wall requiring local correction is present.
    If it is, a further measurement is performed to determine whether the correction operation $U^c_k$ should be applied on site $k=j$ or $j+2$. 
    Which site is corrected depends both on the measurement outcome and on the target state, as illustrated in the table at the bottom.
    }
    \label{fig:control}
\end{figure}

With the chaotic part of the circuit and its unstable fixed point structure in hand, we are now in a position to build a control protocol to push the dynamics onto an unstable fixed point.
The control proceeds in two steps: first, determine the point on the orbit that is closest to the system's current state; second, apply a feedback operation that pushes the system towards that point.
The structure of the control protocol is depicted schematically within the blue box in Fig.~\ref{fig:full-circuit} and described in more detail in Fig.~\ref{fig:control} and in the text below.

To determine the closest point on the vacuum orbit~\eqref{eq: vacuum}, it is convenient to define the sublattice magnetization operators
\begin{align}
    M_{\rm even} = \sum_{j\text{ even}} Z_j,\indent M_{\rm odd} = \sum_{j\text{ odd}} Z_j.
\end{align}
The CB states $\ket{0000\dots}$, $\ket{0101\dots}$, and $\ket{1010\dots}$ on the vacuum orbit are the minimum-eigenvalue eigenvectors of the linear combinations $M_{\rm even}+M_{\rm odd}$, $M_{\rm odd} - M_{\rm even}$, and $M_{\rm even}-M_{\rm odd}$, respectively (all three with eigenvalue $-L$).
We can therefore measure the distance of a quantum state $\ket{\psi}$ to each of these three CB states by performing a quantum measurement of $M_{\rm even}$ and $M_{\rm odd}$.
In particular, for a particular measurement outcome $m_{\rm even(odd)}$, we can define the corresponding eigenspace
\begin{align}
    \mathcal H_{m_{\rm even(odd)}}= \mspan\{\ket{b} \mid \braket{b|M_{\rm even(odd)}|b}=m_{\rm even(odd)}\}
\end{align}
consisting of all CB states $\ket{b}$ with that quantum number.
The Born probability of that measurement outcome is given by $\braket{\psi|P_{m_{\rm even(odd)}}|\psi}$, where
\begin{align}
    P_{m_{\rm even(odd)}}=\sum_{\ket b\in\mathcal H_{m_{\rm even(odd)}}}\ket{b}\bra{b}
\end{align}
is the projector onto the corresponding eigenspace.
Thus the action of the measurement on the state is the non-linear projective process
\begin{align}
    |\psi\rangle \rightarrow \frac{P_{m_{\rm even(odd)}}|\psi\rangle}{\sqrt{\langle \psi | P_{m_{\rm even(odd)}}|\psi\rangle }}.
\end{align}
Note that this measurement does not collapse $\ket{\psi}$ to a single CB state.
In fact, the subspace $\mathcal H_{m_{\rm even(odd)}}$ contains exponentially many CB states and therefore can support volume-law entanglement.
For this reason, we refer to this type of measurement, which reveals global information about the state without fully collapsing it to a product state, as a collective measurement. 
Such a measurement can be implemented experimentally in a qubit-only setup by computing the magnetization onto a register of $O(\log_2 L)$ ancilla qubits~\cite{McArdle19,Wang21,Botelho22,Piroli24}, or in a circuit/cavity QED setup by dispersively coupling the qubits in each sublattice to a separate cavity mode and measuring the frequency shift of either mode. 
We illustrate one explicit way to perform this measurement with ancillary qubits in Appendix~\ref{sec:Magnetization_Measurement}.
Such collective measurements have been achieved (mid-circuit) in arrays of tweezer-trapped Rydberg atoms coupled to cavities~\cite{Deist22,Yan23}.

The first step in the control protocol is to perform collective measurements of $M_{\rm even}$, followed by $M_{\rm odd}$. Given the measurement outcomes $m_{\rm even}$ and $m_{\rm odd}$, we define $m_*$ to be the minimum of $m_{\rm even}+m_{\rm odd}$, $m_{\rm even}-m_{\rm odd}$, and $m_{\rm odd}-m_{\rm even}$. Depending on which of these three values $m_*$ takes, the target state of the control protocol is selected to be $\ket{0000\dots}$, $\ket{1010\dots}$, or $\ket{0101\dots}$, respectively [see table in Fig.~\ref{fig:control}(a)].

The second step in the control is a local feedback operation applied with probability $q$ on each site (see purple box in Fig.~\ref{fig:full-circuit}).
Its specific form depends on the target state determined in the first step, and its aim is to locally correct sublattice domain walls [see Fig.~\ref{fig:control}(b) for a flowchart].
In all three possible cases, it begins with a measurement of the two-body operator $Z_{j}Z_{j+2})$ on site $j$.
If the outcome is $+1$, there is no domain wall and no further correction is necessary.
If the outcome is $-1$, indicating that a domain wall is present, the correction subroutine proceeds by measuring $Z_j$.
Combined with the domain-wall measurement, this yields complete knowledge of the state on sites $j$ and $j+2$.
Depending on which of the three vacuum orbit configurations is ``closest," a unitary correction operation
\begin{align}
\label{eq:correct}
U_k^c=e^{-i\frac{\pi}{2}(PXP)_k},
\end{align}
is applied on site $k=j$ or $j+2$.
$U^c_k$ flips qubit $k$ provided the flip does not violate the Fibonacci constraint described below Eq.~\eqref{eq: UPXP-def}.
The table at the bottom of Fig.~\ref{fig:control}(b) shows whether qubit $j$ or $j+2$ is flipped, depending on the target state and the outcome of the $Z_j$ measurement.
In Appendix~\ref{sec: General Control}, we discuss how this control protocol can be generalized to other Hilbert space orbits with finite recurrence times, of the sort discussed in Ref.~\cite{Iadecola20}.

Both the collective measurements and the resulting feedback are crucial for directing the system to one of the three vacuum states. 
The collective measurements find the ``closest'' orbit, and the feedback adapts based on the measurement result to direct the system closer to the target state.

The strength of the control is quantified by two parameters: the control rate $p$ that sets the probability with which control is applied at a given time step, and the probability $q$ that sets the fraction of sites to which the local feedback operation is applied. 
The addition of the ``spatial" probability $q$ is particularly important in that it represents a step beyond the probabilistic control literature~\cite{antoniou_probabilistic_1996,antoniou_probabilistic_1997,antoniou_absolute_1998,antoniou_probabilistic_2000} that has so far focused on single-body chaotic dynamics~\footnote{In both cases, $p$ and $q$ are classical probabilities implementable, for instance, with a pseudo-random number generator on a classical computer.}.
In generalizing to the many-body setting, we allow for the possibility that only a finite fraction of sites are subjected to control (beyond the initial step in which some global information about the system's state is measured).

Examples of individual dynamical trajectories under the full stochastic protocol including both chaotic and control circuits are shown for $p=0.5$ and $0.6$ in Fig.~\ref{fig:example-evols}(c,d) and (e,f), respectively.
The system size $L=24$ and local correction probability $q=0.2$ are fixed in all four panels.
Two values of the parameter $\theta$ in Eq.~\eqref{eq: U_sigma} are considered: $\theta_Q=\pi/3$ [Fig.~\ref{fig:example-evols}(c,e)] and $\theta_C=\pi/2$ [Fig.~\ref{fig:example-evols}(d,f)].
The former is representative of the fully quantum dynamics, while the latter is representative of the classical limit, which we discuss below.

\subsection{Classical Limit}
\label{sec: Classical Limit}

Both the chaotic circuit and the control protocol have a numerically tractable classical limit that is useful in our analysis of the control transition.
In this limit, the dynamics of an initial CB state does not lead to superposition and entanglement.
Rather, the system is always in a CB state $\ket{b}$, corresponding to a bitstring $b$.
At each time step, the CB state $\ket b \mapsto \ket{b'}$ up to a global phase. 
Thus, to keep track of the dynamics of an initial CB state and the evolution of observables that are diagonal in the CB [like the control order parameter $H_{\rm ZZ}$, Eq.~\eqref{eq:HZZ}], it is sufficient to follow the dynamics of the bitstring $\dots\mapsto b\mapsto b'\mapsto\dots$.
The dynamics in the classical limit is therefore equivalent to a CA.
This is clearly visible in Fig.~\ref{fig:example-evols}(b,d,f) where the system always has expectation value $\braket{Z_j} = \pm 1$ on each site and time, indicating that the system remains in a CB state throughout the dynamics.

To see how this limit arises, we first consider the chaotic circuit defined in Sec.~\ref{sec: Circuit}. The first part of the circuit, $U_{\rm PXP}$, already maps CB states to CB states. 
When acting on a CB state, each local gate $e^{-i\frac{\pi}{2}(PXP)_j}$ is equivalent (up to a global phase) to a Toffoli gate with qubit $j$ as the target and its nearest neighbors as the controls:
\begin{align}
    e^{-i\frac{\pi}{2}(PXP)_2}\ket{000} & = \ket{010}, & e^{-i\frac{\pi}{2}(PXP)_2}\ket{010} & = \ket{000},
\end{align}
and the six other CB states are acted upon as the identity.
The integrability breaking circuit $U_\sigma(\theta)$ generates CA dynamics from CB initial states upon setting $\theta=\pi/2$, so that a full (conditional) flip-flop $01\leftrightarrow 10$ is performed between qubits $j$ and $j+1$.
Note that, because this circuit is unitary, the dynamics of CB states is described by a reversible CA.

The control protocol automatically generates CA dynamics from CB states.
Since the observables measured during this protocol are all diagonal in the CB, an initial CB state will never leave the CB.
Moreover, these measurements become ``classical" in nature: there is no backaction on the quantum state as there are no superpositions to collapse.
Thus the measurements are essentially reading off classical information from the system's state in this limit.
For example, the quantity $m_*$ computed from the sublattice magnetization measurements is essentially the minimum Hamming distance between the current bitstrings and the three possible target bitstrings.
The local unitary correction operation $U^c_k$ [Eq.~\eqref{eq:correct}] also acts as a Toffoli gate on CB states.
The combined action of these processes amounts to an \textit{irreversible} CA because it is not one-to-one.
The local correction procedure is designed to bring the three-site region $\{j,j+1,j+2\}$ into the form $000$, $101$, or $010$ depending on the target state, regardless of the initial state of that region.

\section{Control Transition}
\label{sec: Control Transition}

We now consider the existence of a control transition in this model, before turning our focus in Sec.~\ref{sec: Entanglement Transition} to the interplay of control and entanglement properties.
Previous work on $\beta$-adic R\'enyi circuits~\cite{Iadecola2023,LeMaire24,Allocca24,Pan24} suggests that properties of the control transition are primarily governed by classical physics.
For this reason, we first consider the automaton limit of our model [$\theta=\pi/2$ in Eq.~\eqref{eq: U_sigma}], whose classical simulability gives access to larger system sizes and better estimates of critical properties that can serve as a useful guide when analyzing the quantum limit.
In our numerical simulations, we set $q=0.2$ and measure the control order parameter $H_{ZZ}$ after evolving the system for $L$ time steps from an initial CB state.
(We defer discussion of the $q$-dependence of our results to Sec.~\ref{sec: Phase}.)
Results are averaged over $10^4$ samples comprising randomly chosen initial CB states, circuit realizations, and quantum measurement outcomes (sampled according to their Born probabilities).
Each sample yields an expectation value of the control order parameter that we denote 
$\braket{H_{ZZ}(t)}=\langle \psi(t) |H_{ZZ} | \psi(t) \rangle$.
In the classical limit, $\braket{H_{ZZ}(t)}$ takes discrete values between $0$ and $1$ in steps of $1/L$ since the system is always in a CB state; in the quantum limit, the system can be in a superposition of CB states and $\braket{H_{ZZ}(t)}$ becomes a continuous variable.
The sample average $\overline{\braket{H_{ZZ}(t)}}$, along with error bars denoting the standard error of the mean, is the object of our analysis below.

A universal (classical and quantum) feature of the model is that, at $p=0$, we can take the thermodynamic limit of the order parameter to determine that it saturates to
\begin{equation}
\overline{\braket{H_{ZZ}(t\rightarrow\infty)}}_{p=0}=3/\sqrt{5}-1\approx 0.34164.
\label{eqn:Mz}
\end{equation}
This number is obtained by analytically computing the trace of $H_{ZZ}$ over the Fibonacci Hilbert space; the fact that $\overline{\braket{H_{ZZ}(t\rightarrow\infty)}}_{p=0}<1/2$ is a consequence of the Fibonacci constraint.
At small $p>0$, we find that the order parameter at time $t=L$ takes a value below $\overline{\braket{H_{ZZ}(t\rightarrow\infty)}}_{p=0}$ but remains nonzero.
At large $p$, the order parameter at time $t=L$ approaches $0$ as a function of system size.
Below, we show that these regimes are separated by a continuous phase transition in both the classical and quantum limits of the dynamics, with critical properties broadly consistent in the two limits.

Before proceeding, we offer a remark about finite-size effects. 
The control protocol defined in Sec.~\ref{sec: Control} targets the vacuum orbit Eq.~\eqref{eq: vacuum}, which is invariant under the chaotic circuit defined in Sec.~\ref{sec: Circuit}.
If the control protocol successfully drives the system onto the orbit, it can never escape.
In a finite-size system, even at low control rate there is a finite probability that the system becomes trapped on the orbit.
This probability is exponentially small in the system size $L$, but this means that such rare events can occur on timescales $\sim e^{O(L)}$.
At large system sizes like those accessible in the classical limit, this is not an issue.
However, at small system sizes like those accessible in the quantum limit, this finite-size effect manifests as a late-time decay of the control order parameter, even in the chaotic phase where one would expect it to saturate to a time-independent value.
We discuss the decay further in Appendix~\ref{sec: Sticky Orbits} and provide a mechanism to remove it if desired.
Nevertheless, we will see that it is still possible to perform a consistent finite-size scaling analysis in the presence of this decay~\footnote{We note that this effect is also present in studies of absorbing state transitions in monitored quantum dynamics~\cite{Buchhold2022,Ravindranath22,ODea22}, but was not present in studies of probabilistic control in $\beta$-adic R\'enyi circuits~\cite{Iadecola2023,LeMaire24,Allocca24,Pan24}, where a random perturbation was used to remove the orbits while preserving the control transition.}.

\subsection{Classical Control Transition}

\begin{figure}[tb!]
    \centering
    \includegraphics[width=\columnwidth]{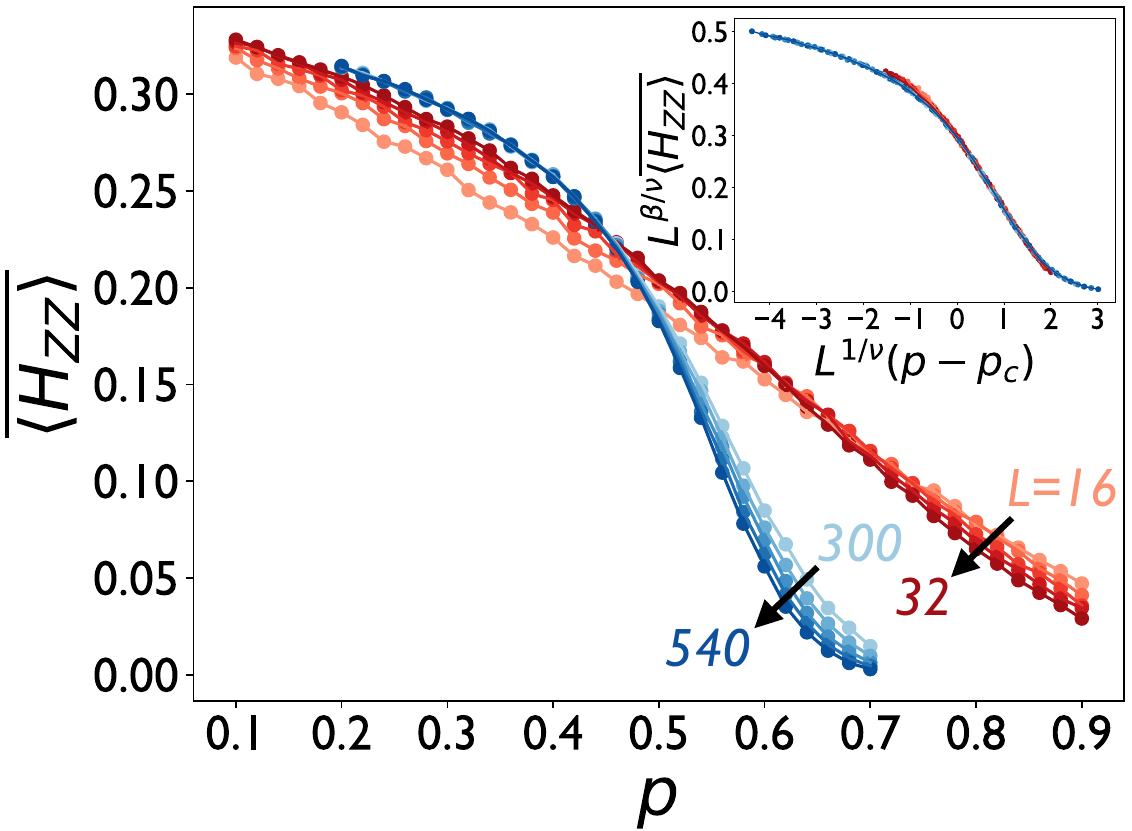}
    \caption{
    \textbf{Order parameter (classical model)}. 
    Shown at time $t=L^{0.86}$ as a function of control rate $p$ in the classical limit of the model for both large (blue) and small (red) system sizes.
    Here and in all subsequent figures, error bars indicating the standard error of the mean over $10^4$ samples are present, but are smaller than the plot markers.
    Inset: Finite-size scaling collapse of the large-system data indicates $p_c=0.49(1)$, $\beta = 0.17(2)$, and $\nu=2.3(2)$. 
    Small-system data are collapsed assuming a scaling function of the form $L^{-\beta/\nu}f[L^{1/\nu}(p-p_c)+AL^{-\alpha}]$ with $p_c$, $\beta$, and $\nu$ fixed by the large-system data collapse. This yields $A\approx 9.61$ and $\alpha\approx 1.08$, which quantify the finite-size effects present at system sizes accessible to the quantum numerics. 
    }
    \label{fig:hzz}
\end{figure}

\begin{figure}[tb!]
    \centering
    \includegraphics[width=\columnwidth]{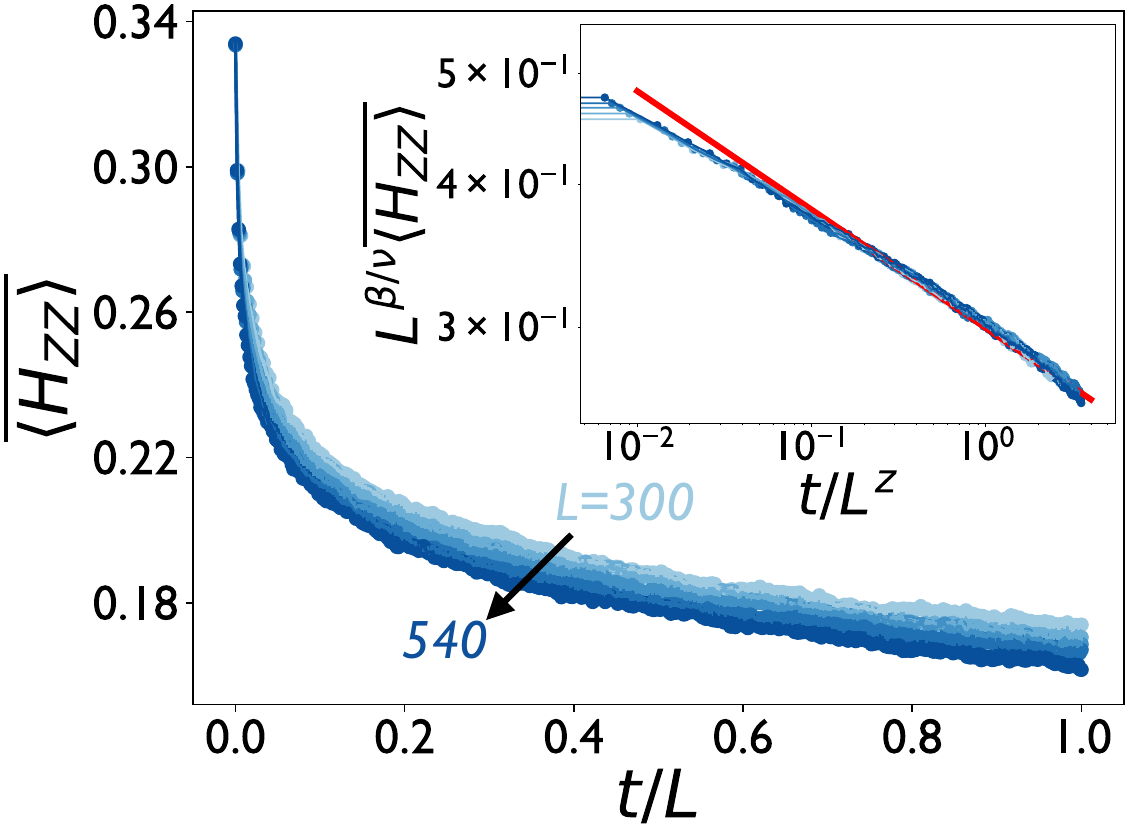}
    \caption{\textbf{Dynamics of the order parameter (classical model).} Realization-averaged order parameter at $p=0.5$, near the control transition. Inset: Finite-size scaling collapse of the order-parameter dynamics yields a dynamical exponent $z=0.8(1)$. The red line shows a fit to a power-law decay $\sim 1/t^{0.104}$ which is close to the expected decay exponent $\beta/(\nu z)\approx 0.093$.}
    \label{fig:hzz-evol}
\end{figure}

In Fig.~\ref{fig:hzz}, we plot the sample-averaged order parameter $\overline{\braket{H_{ZZ}(t=L^{0.86})}}$~\footnote{Note that, since $z$ is not generically an integer, evaluating observables at time $t=c\, L^z$ for some $O(1)$ constant $c$ requires obtaining $\overline{S_{\rm anc}}$ at noninteger times.
In this work we use a linear interpolation to estimate the values of observables at $t=c\, L^z$ and to determine the associated error bars from those of the neighboring data points.} for a range of $p$ and two families of system sizes: $L=300,\dots,540$ (blue points) and $L=16,\dots,32$ (red points).
The exponent $0.86$, which yields the best scaling collapse under the protocol described in the next paragraph, is an estimate of the dynamical critical exponent $z$.
From these data, we extract an estimate of the critical point $p_c$.
The dynamics near $p_c$, shown in Fig.~\ref{fig:hzz-evol}, are then analyzed to yield a refined estimate $z=0.8(1)$ which is within error bars of the value used to obtain Fig.~\ref{fig:hzz}.
The large-system-size data will be used to extract the location and critical exponents of the control transition.
The small-system-size data represent system sizes accessible to our fully quantum simulations that will be reported in the next subsection.
We first describe the transition from the perspective of the large-system data before returning to the small-system data to analyze the finite-size effects that will be important when considering the quantum data.

To extract the critical point $p_c$, we perform finite-size scaling collapse assuming a scaling ansatz $\overline{\braket{H_{ZZ}(t=L^z)}}=L^{-\beta/\nu}f[L^{1/\nu}(p-p_c)]$.
This corresponds to a second-order phase transition where the order parameter vanishes near the transition as 
$\overline{\braket{H_{ZZ}(t=L^z)}}_{p\sim p_c}\sim |p-p_c|^\beta$ 
(in the thermodynamic limit)
and the correlation length diverges as $\xi\sim |p-p_c|^{-\nu}$.
The scaling collapse (like all others reported in this paper) is performed using a standard $\chi^2$ analysis~\cite{Kawashima93} that uses a numerical least-squares opimization to collapse the curves.
This analysis yields estimates of the transition point $p_c=0.49(1)$, the order-parameter critical exponent $\beta=0.17(2)$, and the correlation-length critical exponent $\nu=2.3(2)$ (see inset for collapsed curves).
Error bars on these quantities are obtained as in Ref.~\cite{Zabalo20}, namely by examining where in the three-dimensional parameter space the $\chi^2$ cost function exceeds the minimum value by 30\%.

We now characterize the finite-size effects at the control transition by considering the small-size data in Fig.~\ref{fig:hzz} (red data points), which corresponds to the system sizes for our simulations of the quantum limit of the model described in the next section.
The order parameter for the small-system data shows a finite-size crossing near $p=0.7$, which naively would suggest that the order parameter remains finite at the transition and that the transition occurs at a value significantly larger than the one obtained from the large-system results.
However, using the large-system results as a prior allows one to extract the critical properties from these data as well.
In particular, we can analyze the small-system data assuming a scaling function of the form 
\begin{equation}
   \overline{\braket{H_{ZZ}(t= L^z)}}_{p>p_c}\!\!\sim\! 
   \frac{1}{L^{\beta/\nu}}
   f\!\left[\!L^{1/\nu}(p-p_c)\!+\!A_CL^{-\alpha}\!\right]
   \label{eqn:Mzscaling}
\end{equation}
with the values of $\beta,\nu,$ and $p_c$ fixed to those extracted from the large-system data.
$A_C$ and $\alpha$, which characterize the finite-size effects of the classical model, are then taken as fitting parameters for a separate $\chi^2$ analysis.
Their optimal values $A_C\approx 9.61$ and $\alpha\approx 1.08$ collapse the small- and large-system data, as shown in the inset of Fig.~\ref{fig:hzz}.
We note that the finite-size correction to the scaling variable, $A_CL^{-\alpha}$, takes values between $\sim0.48$ ($L=16$) and $\sim 0.23$ ($L=32$) for the small-system data (compared to $\lesssim0.02$ for the large-system data), indicating strong finite-size effects at the system sizes accessible to simulations of the quantum limit of the model.

The dynamical exponent $z$ of the control 
can be deduced from the time evolution of the order parameter near the transition.
In Fig.~\ref{fig:hzz-evol}, we show the dynamics of the order parameter for the large-system data at $p=0.5$.
To determine the dynamical exponent, we collapse these data to a function of the form $L^{-\beta/\nu}f(t/L^z)$ with $\beta/\nu$ fixed by the analysis in the previous paragraph. This leads to an estimated dynamical exponent $z=0.8(1)$ that produces the collapse shown in the inset.
As a further self-consistency check, we note that combining the spacetime scaling $t\sim\xi^z$ with the scaling forms $\xi\sim|p-p_c|^{-\nu}$ and $\overline{\braket{H_{ZZ}(t=L^z)}}_{p\sim p_c}\sim|p-p_c|^\beta$, we are led to the prediction that the order parameter should decay with time as a power law $\sim t^{\beta/(\nu z)}\approx t^{-0.093}$ near the transition.
The inset of Fig.~\ref{fig:hzz-evol} shows the collapsed curves on a log-log scale, with a red line indicating a best-fit power law obtained from the $L=540$ data.
The best-fit power law $\sim t^{-0.104}$ is close to the value expected based on the above scaling argument.

It is notable that a dynamical exponent $z<1$ is obtained from this analysis.
An exponent $z=1$ is expected in transitions governed by a conformal field theory, as in the space-time random MIPT~\cite{Zabalo20}. 
In adaptive circuits, absorbing-state transitions belonging to the directed-percolation and parity-conserving universality classes have $z\approx1.581$ and $1.744$, respectively~\cite{Hinrichsen2000,ODea22,Ravindranath22}. (Further, recent work shows a family that interpolates between these classes \cite{ODeaKhemani2024}.)
An exponent $z=2$, associated with diffusively spreading critical correlations governed by a random walk, has been observed in the Bernoulli map with both local~\cite{LeMaire24,Allocca24,Pan24} and global feedback~\cite{Iadecola2023,Pan24}.
Infinite randomness~\cite{Zabalo23} and quasiperiodic~\cite{Shkolnik23} fixed points with $z>1$ have also been identified in stabilizer circuits with static but spatially varying measurement probabilities.
On the other hand, dynamical exponents $z<1$ have been found to occur in MIPTs with ``power-law interacting" entangling gates~\cite{Block22} and at equilibrium phase transitions with sufficiently long-range power-law interactions~\cite{Maghrebi16} and are associated with superballistic spreading of correlations and sublinear power-law lightcones~\cite{Chen19}.
We attribute this feature of the transition to the sublattice magnetization measurements, which are themselves nonlocal, albeit spatially uniform.

\subsection{Quantum Control Transition}
\label{sec: Quantum Control Transition}

Having established the existence of a control transition in the classical limit of the model, we now ask how the situation changes when the chaotic dynamics becomes intrinsically quantum. 
To do so, we change the parameter $\theta$ in Eq.~\eqref{eq: U_sigma} from $\pi/2$ to $\pi/3$, so that the dynamics now generates superpositions when acting on CB states. 
At the same time, the special structure of the unitary circuit $U_\sigma$ ensures that the vacuum orbit \eqref{eq: vacuum} remains an unstable periodic orbit under the quantum dynamics. 

\begin{figure}[tb!]
    \centering
    \includegraphics[width=\columnwidth]{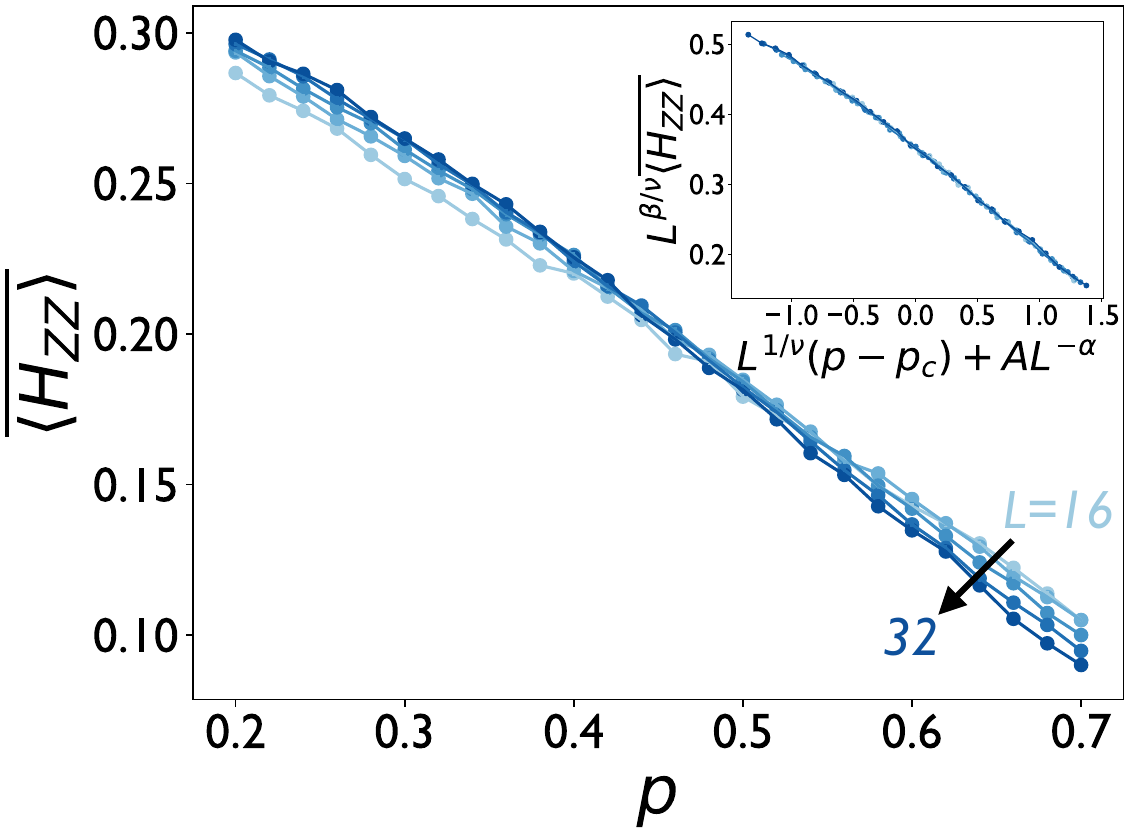}
    \caption{
    \textbf{Order parameter (quantum model).} Shown at time $t=L^{0.85}$ as a function of control rate $p$ for the quantum model.
    Inset: Finite-size scaling collapse indicates $p_c=0.46(5)$, $\beta = 0.32(4)$, $\nu=2.05(25)$. $A\approx 28.2$ and $\alpha\approx 1.54$ provide the leading corrections to finite-size scaling.}
    \label{fig:quantum-hzz}
\end{figure}

\begin{figure}[tb!]
    \centering
    \includegraphics[width=\columnwidth]{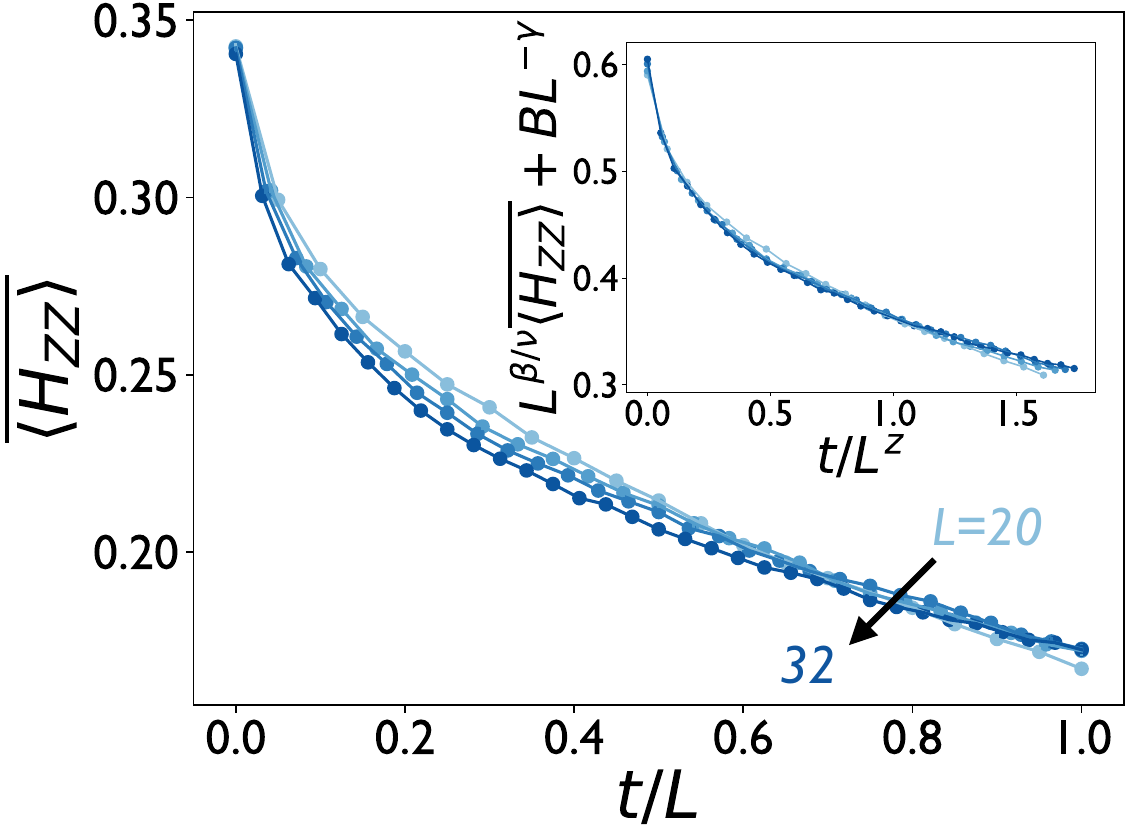}
    \caption{
    \textbf{Dynamics of the order parameter (quantum model).} Shown at $p=0.46$, near the control transition.
    The inset shows a scaling collapse using the values of $\beta$ and $\nu$ extracted from Fig.~\ref{fig:quantum-hzz}.
    The collapse yields an estimated dynamical exponent $z=0.85(5)$.
    $B\approx8.06$ and $\gamma \approx 1.76$ provide the leading corrections to finite-size scaling.
    }
    \label{fig:quantum-hzz-evol}
\end{figure}

Our analysis of the quantum control transition follows that of the classical transition presented in the previous subsection.
In Fig.~\ref{fig:quantum-hzz}, we plot the order parameter $\overline{\braket{H_{ZZ}(t=L^{0.85})}}$ averaged over $10^4$ circuit realizations for system sizes $L=16,\dots,32$ and $p=0.2,\dots,0.7$.
The exponent $0.85$ is chosen to achieve the best collapse, and a direct analysis of the dynamics near the critical point (Fig.~\ref{fig:quantum-hzz-evol}) yields an estimated dynamical exponent $z=0.85(5)$ (see discussion below).
The results are qualitatively similar to the small-system results for the classical model shown in Fig.~\ref{fig:hzz}: the order parameter grows with $L$ at small $p$ and shrinks with $L$ at large $p$, with a finite-size crossing near $p=0.5$.
However, our analysis of the small-system classical data demonstrated that this crossing was a red herring, with the large-system data indicating a transition well below the small-system crossing point with the order parameter approaching zero rather than remaining finite.
With this in mind, we perform a finite-size scaling analysis of the quantum data assuming a critical scaling form given in Eq.~\eqref{eqn:Mzscaling} for the order parameter data, which can take into account the expected substantial finite-size effects.
The data collapse obtained from a 5-parameter $\chi^2$ analysis is shown in the inset of Fig.~\ref{fig:quantum-hzz}.
The analysis yields an estimate of the transition location $p_c=0.46(5)$, order-parameter exponent $\beta=0.32(4)$, and correlation-length exponent $\nu=2.05(25)$.
Finite-size effects are accounted for by $A\approx 28.2$ and $\alpha\approx1.54$, leading to finite-size corrections ranging from $AL^{-\alpha}\approx 0.39$ ($L=16$) to $0.14$ ($L=32$), which is on the same order as those obtained for the small-system classical data.
The transition location $p_c$ and correlation-length exponent $\nu$ agree within error bars with those obtained for the classical transition, while the order parameter exponent $\beta$ is substantially larger (by about a factor of 2) for the quantum transition.

To extract the dynamical critical exponent $z$, in Fig.~\ref{fig:quantum-hzz-evol} we show the dynamics of the order parameter for various system sizes at $p=0.46$, near the control transition.
Owing to the substantial finite size effects observed in the late-time value of the order parameter considered above, we allow for a system-size dependent but time independent shift of the order parameter dynamics such that $L^{\beta/\nu}\overline{\braket{H_{ZZ}(t)}}+BL^{-\gamma} = f(t/L^z)$ is our scaling function [which can be obtained by Taylor expanding Eq.~\eqref{eqn:Mzscaling}].
Using the value $\beta/\nu=0.157$ obtained from Fig.~\ref{fig:quantum-hzz}, we perform a 3-parameter $\chi^2$ analysis that yields $z=0.85(5)$ (see inset).
Finite-size effects are accounted for by $B\approx8.06$ and $\gamma \approx 1.76$, corresponding to finite size corrections ranging from $BL^{-\gamma}\approx0.04$ ($L=20$) to $0.02$ ($L=32$). 
While these corrections are about an order of magnitude smaller than those obtained for the late-time order parameter data, they should be compared to the scale of the vertical axis in Fig.~\ref{fig:quantum-hzz-evol}, which is also an order of magnitude smaller than that of the horizontal axes in Figs.~\ref{fig:quantum-hzz} and \ref{fig:hzz}.
Thus the finite size corrections to our scaling analyses are broadly consistent.
Moreover, the value $z=0.85(5)$ matches the value $z=0.8(1)$ obtained for the classical transition in Fig.~\ref{fig:hzz-evol}.
This further indicates that the ``long-range" nature of the classical transition persists in the quantum limit.

\section{Entanglement Transition}
\label{sec: Entanglement Transition}

Promoting the CA dynamics to be fully quantum has several important implications.
First, the unitary part of the dynamics described in Sec.~\ref{sec: Circuit} now also leads to the generation of entanglement as a function of time.
Second, the measurements involved in the control protocol now entail a nontrivial backaction on the quantum state.
This backaction tends to remove entanglement and competes with the entanglement generation due to the unitary part of the dynamics.
This leads to the possibility of an entanglement phase transition as well as a control transition.
Whether these two transitions coincide or occur separately is a delicate question that depends on the locality of the control protocol~\cite{SierantTurkeshi2023,LeMaire24,Pan24}. 

In this section, we investigate the entanglement transition through the lens of three entanglement measures: the half-cut von-Neumann entanglement entropy $S_A$ (with subsystem $A$ defined to consist of the left half-chain), the tripartite mutual information $\mathcal I_3$ (between three equal-size subregions) associated with this entanglement entropy, and the von-Neumann entanglement entropy $S_{\rm anc}$ of an ancilla qubit that is initially maximally entangled with the system.
Of these, we use $\mathcal I_3$ and $S_{\rm anc}$ to provide independent estimates of the critical properties of the entanglement transition.
We find that the transition points and critical exponents extracted from these entanglement measures agree within numerical resolution with one another and with those extracted for the control transition in Sec.~\ref{sec: Control Transition}. 
The fact that the locations and critical exponents agree among the various quantities provides strong evidence that the control and entanglement transitions coincide.
However, we are not able to definitively rule out the possibility that the transitions are weakly split but have similar criticality.

As with the late-time decay of the order parameter discussed in Sec.~\ref{sec: Control Transition}, the entanglement decays in both phases due to finite size effects.
We account for this in entanglement measures with the schematic form of decay 
\begin{align}
\label{eq:SA-decay}
    \overline{S_{A}}(t,p,L) \sim \exp[-\Gamma(p,L)\, t/L] S_{\infty}(p,L),
\end{align}
where $\Gamma(p,L)$ is a decay rate and the scaling of $S_{\infty}(p,L)$ encodes the properties of the entanglement phase.
Further details on this decay, and a discussion of how to remove it entirely, are discussed in Appendix~\ref{sec: Sticky Orbits}. 
We expect that these signatures are present in previously studied absorbing-state transitions~\cite{Buchhold2022,ODea22,Ravindranath22}, although the effects may be small if large systems are being studied with, e.g., Clifford or free-fermion methods. 
It is also possible that the probability of disentangling the system by becoming accidentally trapped on the absorbing state is larger in the presence of global measurements like those employed in our control protocol.
Quantifying these differences could help elucidate controllability within the chaotic (uncontrolled) regime; we leave this for future work.

An important observation made in Appendix~\ref{sec: Sticky Orbits} is that the decay rate $\Gamma(p,L)$ is approximately $L$-independent close to the transition and deep in the area-law phase. Thus, evaluating the entanglement entropy at a time $t$ of order $L$ allows us to probe the criticality via the entanglement scaling encoded in $S_\infty(p,L)$ in Eq.~\eqref{eq:SA-decay}, although some residual noise in the data due to these systematics is still to be expected.


\subsection{Entanglement Entropy Scaling}
\label{sec: Entanglement Entropy Scaling}

\begin{figure}[tb!]
    \centering
    \includegraphics[width=\columnwidth]{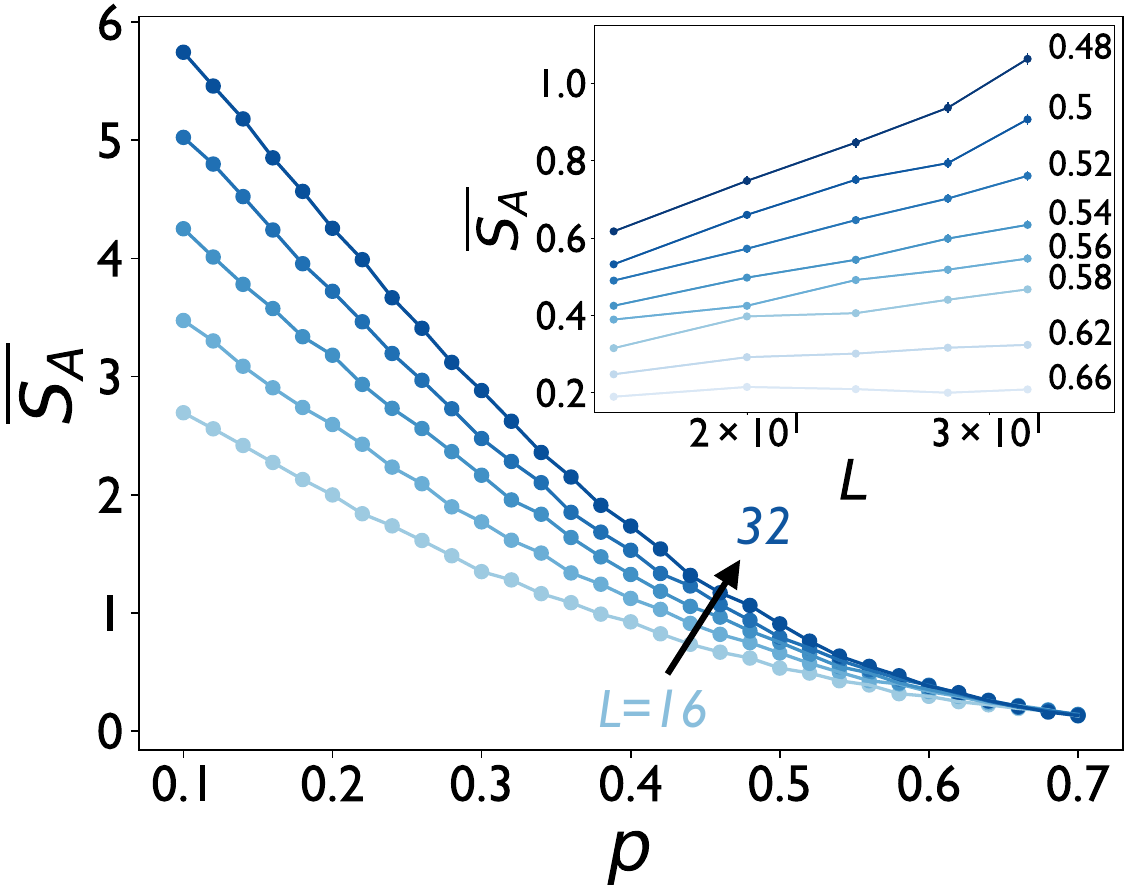}
    \caption{\textbf{Half-cut entropy (quantum model)}. Sample-averaged half-cut entanglement entropy at time $t=L$ as a function of $p$ and several system sizes.
    Inset: System-size dependence of the same data for a range of $p$ values, with a logarithmic scale on the horizontal axis.}
    \label{fig:half-cut-tL}
\end{figure}

The system-size scaling of the late-time entanglement entropy $\overline{S_A(t=L)}$ as a function of $p$ is plotted in Fig.~\ref{fig:half-cut-tL}.
As expected, $\overline{S_A(t=L)}$ grows with $L$ at small $p$ but saturates to a constant at small $p$.
The dependence of $\overline{S_A(t=L)}$ on $L$ for several fixed values of $p$ is shown in the inset, with a logarithmic scale on the $x$ axis.
For intermediate values of $p\sim 0.52$--$0.56$, the growth with $L$ is approximately logarithmic, as observed for example at the MIPT.
However, distinguishing logarithmic from power-law growth with so few values of $L$ is highly nontrivial, and the observation of this behavior over a range of $p$ makes it difficult to pinpoint an entanglement transition from the entanglement entropy itself.

\begin{figure}[tb!]
    \centering
    \includegraphics[width=\columnwidth]{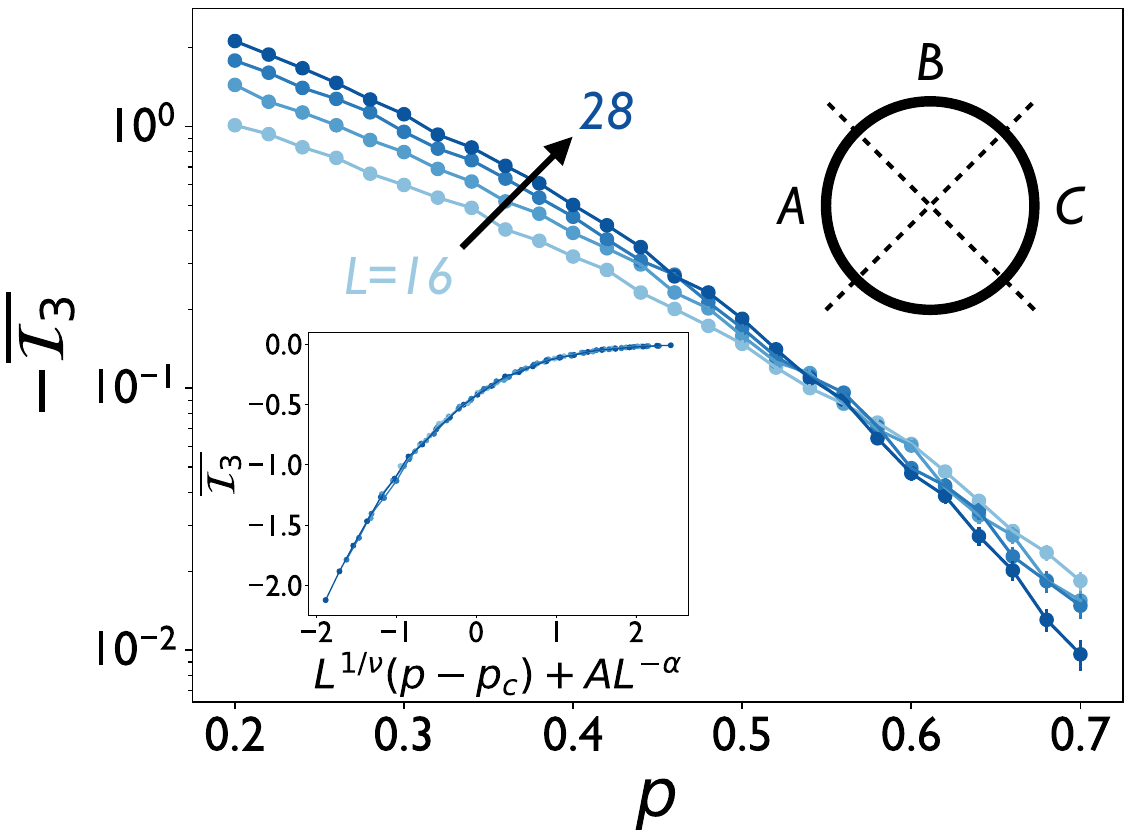}
    \caption{
    \textbf{Tripartite mutual information (quantum model).} TMI as a function of $p$ for several system sizes for subsystems $A,B,C$ arranged as shown in the inset.
    $-\overline{\mathcal I_3}$ is plotted on a logarithmic scale on the vertical axis to improve visibility around $p=0.5$, where finite size crossings occur.
    Inset: Scaling collapse yields
    $p_c=0.44(2)$ and $\nu=1.6(2)$, with $A\approx25.9$ and $\alpha\approx1.40$ providing the leading corrections to finite size scaling.
    \label{fig:tmi}
    }
\end{figure}

Because of these difficulties, which arise also in the MIPT, the tripartite mutual information (TMI), defined as
\begin{align}
\begin{split}
\mathcal I_3 &= S_A + S_B + S_C\\
&\qquad- S_{A\cup B} - S_{A\cup C} - S_{B\cup C} + S_{A\cup B\cup C} 
\end{split}
\end{align}
is a useful quantity to consider~\cite{Zabalo20}.
It is defined with respect to a partition of the 1D system with periodic boundary conditions into four subsystems $A,B,C,$ and their complement, which we each take to consist of $L/4$ sites, as shown in the inset of Fig.~\ref{fig:tmi}.
$S_R$ is the von-Neumann entanglement entropy with respect to region $R$ consisting of one or more of these subsystems.
$\mathcal I_3$ is expected to flow to zero as $L\to\infty$ in an area-law phase, and to $-\infty$ as $L\to\infty$ in a volume-law phase.
In contrast, at a critical point where the entanglement entropy grows logarithmically with the subsystem size, the TMI is expected to take a system-size-independent value, which manifests as a crossing point when plotted as a function of $p$ for various $L$.

Fig.~\ref{fig:tmi} shows the TMI at time $t=L$ as a function of $p$ for $L=16,\dots,28$.
It displays the expected behavior at large and small $p$ and appears to manifest several finite-size crossings between $p\approx 0.45$ and $0.55$, where the largest and smallest pair of consecutive system sizes, respectively, exhibit crossings.
The lack of a single clear crossing is likely due to the finite-size effects discussed in Sec.~\ref{sec: Control Transition} and Appendix~\ref{sec: Sticky Orbits}.
To mitigate these finite-size effects, we perform a scaling collapse using the same ansatz adopted in Fig.~\ref{fig:quantum-hzz}, with the result shown in the inset.
This collapse yields an estimated critical point $p_c=0.44(2)$ and correlation-length exponent $\nu=1.6(2)$, consistent within error bars with the corresponding quantities extracted for the control transition in Fig.~\ref{fig:hzz}.
The best collapse is achieved by applying the finite size scaling ansatz 
\begin{equation}
\overline{\mathcal{I}_3(t/L^z\to\infty)}_{p>p_c}\sim 
g\left[L^{1/\nu}(p-p_c)+A_QL^{-\alpha}\right]
   \label{eqn:I3scaling}
\end{equation}
with finite-size corrections controlled by $A_Q\approx 25.9$ and $\alpha\approx 1.40$, corresponding to finite-size corrections $A_QL^{-\alpha}\approx 0.53$ ($L=16$) to $0.24$ ($L=28$), which are on par with the corrections obtained from the collapses in Figs.~\ref{fig:hzz} and \ref{fig:quantum-hzz}.

\subsection{Purification Transition}
\label{sec: Purification Transition}

The entanglement entropy scaling results described above suggest an entanglement transition that coincides (to within our numerical resolution) with the control transition.
An independent check of these results can be obtained by viewing the entanglement transition as a purification transition~\cite{gullans_dynamical_2020,Gullans20} measured by the time it takes for measurements to purify an initially mixed quantum state.
A useful order parameter for the transition is $S_{\rm anc}$, the von-Neumann entropy of a single ancilla qubit that is initially maximally entangled with the full system on which the adaptive dynamics protocol is being performed.
[In this paper we define $S_{\rm anc}$ in units of $\ln(2)$ for convenience.]
At low measurement rates, the measurements are not performed frequently enough to disentangle the system from the ancilla, so the system is in a ``mixed phase" where $S_{\rm anc}$ approaches $1$ as $L\to\infty$.
At high measurement rates, the system is rapidly disentangled from the ancilla, so the system is in a ``pure phase" where $S_{\rm anc}$ approaches $0$ as $L\to\infty$.
Studies of the MIPT have shown that the mixed and pure phases correspond with the volume- and area-law entanglement phases~\cite{Gullans20}, demonstrating the utility of $S_{\rm anc}$ as a probe of the MIPT.
To measure $S_{\rm anc}$, we prepare the ancilla in a state that is maximally entangled with the full system, namely
\begin{align}
    \ket{\Psi_0} = \frac{1}{\sqrt 2}(\ket{0}_a\ket{\psi_1}+\ket{1}_a\ket{\psi_2}),
\end{align}
where $\ket{0}_a$ and $\ket{1}_a$ are the computational basis of the ancilla and $\ket{\psi_{1,2}}$ are orthogonal volume-law states of the primary system.
In practice, for each circuit realization we choose two states $\ket{\psi_1}$ and $\ket{\psi'_{2}}$ uniformly at random according to the Haar measure on the Fibonacci Hilbert space and then orthogonalize $\ket{\psi'_2}$ against $\ket{\psi_1}$ to obtain $\ket{\psi_2}$~\footnote{Note that two independently sampled random states of $L$ qubits have an overlap that is exponentially small in $L$, so the orthogonalization does not result in a large change to the state $\ket{\psi'_2}$.}.
We then evolve the primary qubit register out to time $t=L$ to obtain the data discussed below.

\begin{figure}[tb!]
    \centering
    \includegraphics[width=\columnwidth]{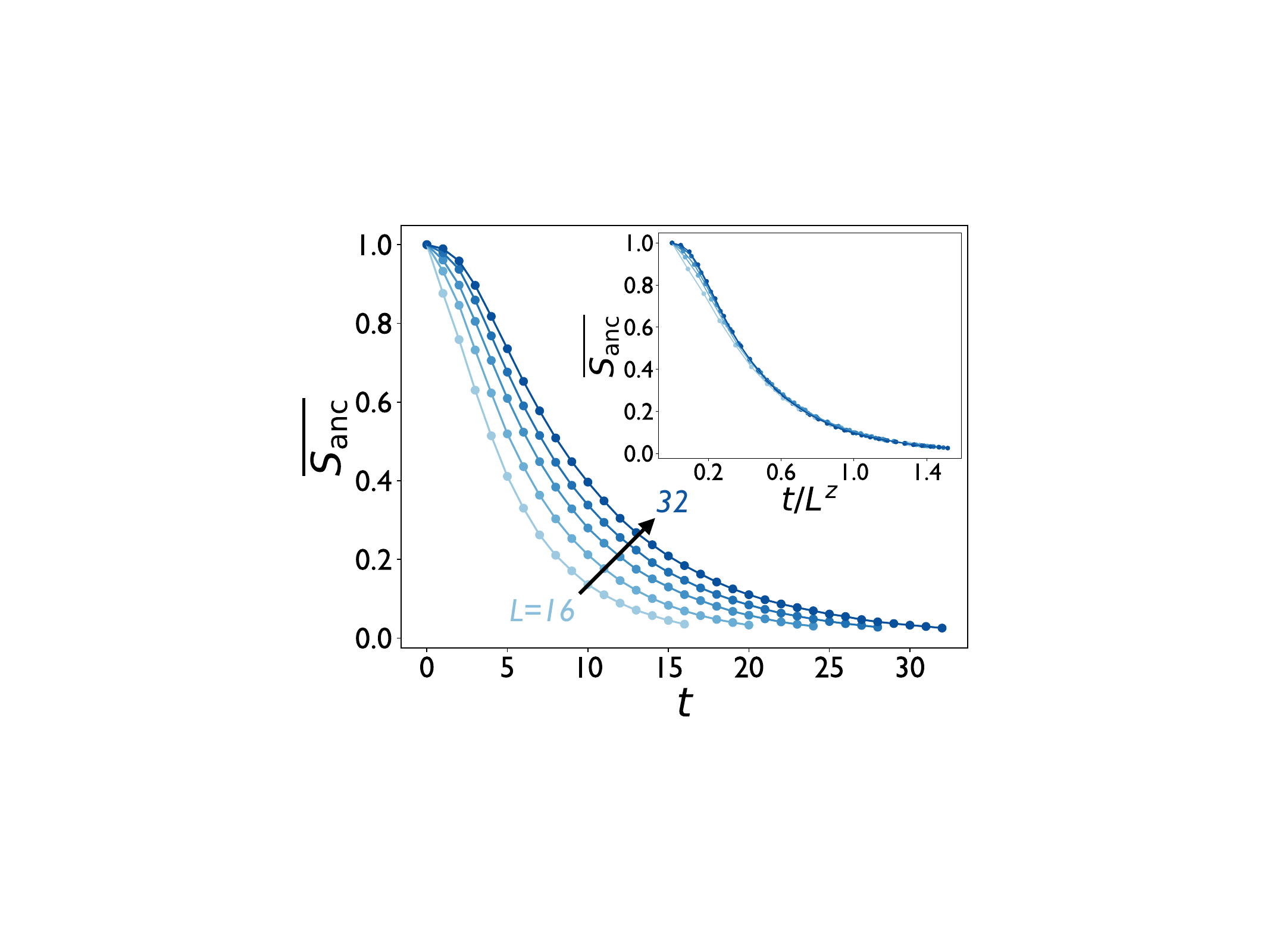}
    \caption{\textbf{Dynamics of ancilla entropy (quantum model).} Time series of the sample-averaged ancilla entanglement entropy in units of $\ln 2$ at $p=0.42$ for several system sizes.
    Inset: A one-parameter scaling collapse yields
    $z=0.88(2)$. 
    }
    \label{fig:ancilla-evol}
\end{figure}

Fig.~\ref{fig:ancilla-evol} shows the dynamics of $\overline{S_{\rm anc}}$ as a function of $t$ for $L=16,\dots,32$ at $p=0.42$.
Regardless of the value of $p$, $\overline{S_{\rm anc}}$ starts at 1 and then decays to $0$ as a function of time.
Near the purification transition $p=p_c$, the ancilla purifies on a timescale $t\sim L^z$, where $z$ is the dynamical exponent of the transition.
Thus, near the transition the time series of $\overline{S_{\rm anc}}$ 
obeys the scaling function 
\begin{equation}
    \overline{S_{\rm anc}}(t,L)_{p=p_c}\sim h(t/L^z)
\end{equation}
and 
we estimate $z$ in the following manner.
At each $p$, we perform a $\chi^2$ analysis to collapse the dynamics of $\overline{S_{\rm anc}}$ for $t>5$ as a function of $t/L^z$ with $z$ taken as the only fitting parameter.
After obtaining the optimal $\chi^2$ value for each $p$, we select the $z$ value yielding the smallest optimal $\chi^2$ and take the corresponding $p$ value as an estimate of the transition.
This analysis yields $z=0.88(2)$, with both $p=0.4$ and $p=0.42$ producing optimal $z$ values within this range.
Remarkably, this is consistent with the dynamical exponent obtained for the classical control transition, see Fig.~\ref{fig:hzz-evol}.

\begin{figure}[tb!]
    \centering
    \includegraphics[width=\columnwidth]{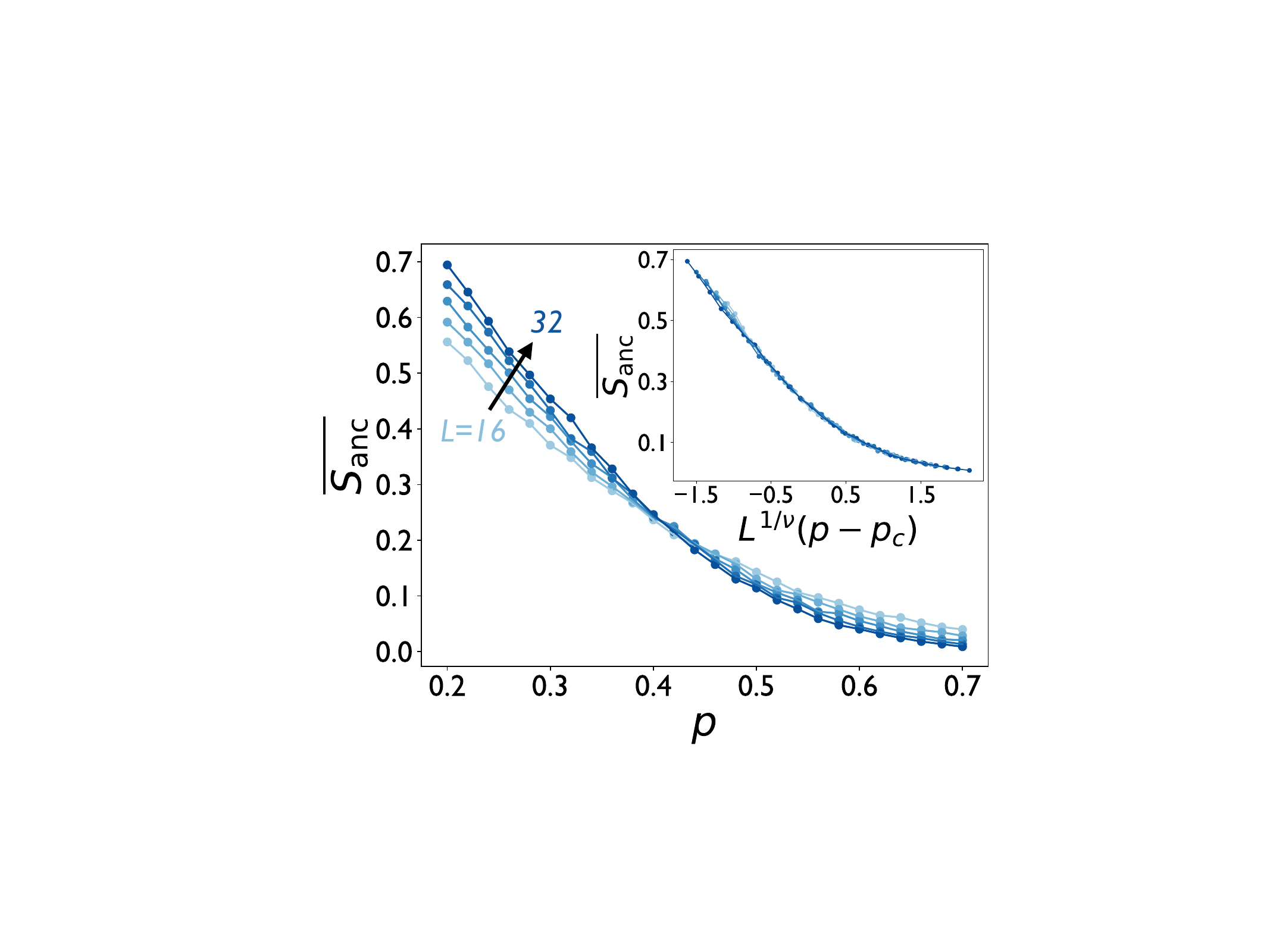}
    \caption{\textbf{Ancilla entropy (quantum model).} Ancilla entanglement entropy in units of $\ln 2$ measured at time $t=0.7L^{0.88}$ for various $p$ and $L$.
    A linear interpolation is used to produce data points and error bars (standard error of the mean) at noninteger values of $t$.
    Inset: Two-parameter scaling collapse yields $p_c=0.41(2)$ and $\nu=1.7(2)$. }
    \label{fig:ancilla}
\end{figure}

To further pinpoint the purification transition and estimate the correlation-length critical exponent $\nu$, we use the system-size scaling of $\overline{S_{\rm anc}}$ evaluated at a time of order $L^z$.
Measuring this quantity as a function of $p$ for various $L$, we expect to see curves for different $L$ cross at $p_c$, with $\overline{S_{\rm anc}}$ increasing (decreasing) as a function of $L$ below (above) this value.
This is shown in Fig.~\ref{fig:ancilla}, which plots $\overline{S_{\rm anc}}$ at time $t=0.7L^{0.88}$ (we confirmed that the results described below do not depend strongly on the choice of the prefactor $0.7$---any prefactor $\gtrsim 0.6$ works well as long as the value of $\overline {S_{\rm anc}}$ is not too small).
A scaling collapse assuming a scaling function $f[L^{1/\nu}(p-p_c)]$ is then performed (see inset), yielding the estimates $p_c=0.41(2)$ and $\nu=1.7(2)$.
These values serve as independent estimates of the location and critical properties of the entanglement transition, and are consistent with those obtained from the TMI data in Fig.~\ref{fig:tmi}.
Notably, however, the finite size effects are much less pronounced for the ancilla entropy, making it perhaps the most reliable witness of the entanglement transition.
In fact, carrying out the collapse with a scaling ansatz of the form \eqref{eqn:I3scaling} yields finite size corrections that are zero to machine precision, giving the same $p_c$, $\nu$, and error bars.

\section{Exploring the Phase Diagram}
\label{sec: Phase}

So far, we have focused on the control and entanglement transitions as a function of the control rate $p$.
However, the properties and locations of the two transitions also depend on the parameter $q$ (fixed at $0.2$ in Secs.~\ref{sec: Control Transition} and \ref{sec: Entanglement Transition}), which sets the fraction of sites on which local domain wall corrections are performed.
Intuitively, we expect $p_c$ to drift upwards with decreasing $q$.
At $q=1$, when all domain walls are measured and corrected, the system immediately disentangles and becomes controlled in an $O(1)$ time, so $p_c=0$.
At $q=0$, when no corrections are performed, the system never becomes controlled and remains volume-law entangled for any $p$, so $p_c\to\infty$.
As $q$ is increased, we expect the dynamical exponent $z$ to flow downwards and the correlation length exponent $\nu$ to flow upwards.
As more domain-wall measurements and corrections are performed, we expect control to occur more rapidly in time, and this should be reflected in the timescale for control at criticality via a decrease in $z$.
Moreover, since the domain-wall measurements and corrections are conditioned on the same collective measurement outcomes, we expect them to generate correlations in space that should be reflected in the divergence of the correlation length by an increase in $\nu$.

\begin{figure}[tb!]
    \centering
    \includegraphics[width=\columnwidth]{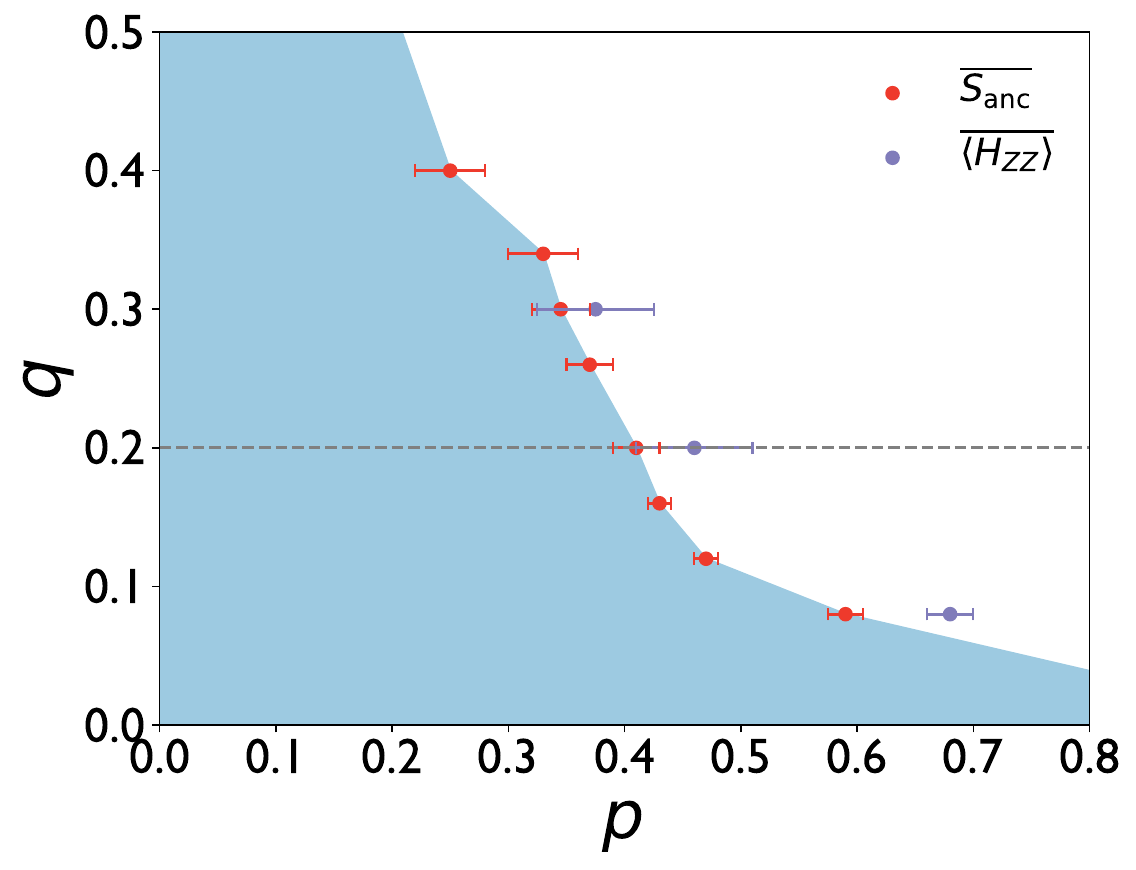}
    \caption{\textbf{Phase diagram (quantum model).} The locations of the entanglement ($\overline{S_{\rm anc}}$) and control ($\overline{\braket{H_{ZZ}}}$) transitions, along with their error bars from $\chi^2$ analysis, are shown in red and purple, respectively. Here, $p$ is the probability that the control circuit is applied at each time step, and $q$ is the fraction of sites undergoing a local domain wall correction when the control is applied. The shaded region represents the volume-law uncontrolled phase. The dashed line at $q=0.2$ represents the line cut considered in  previous sections.}
    \label{fig:phase-diagram}
\end{figure}

\begin{figure}[tb!]
    \centering
    \includegraphics[width=\columnwidth]{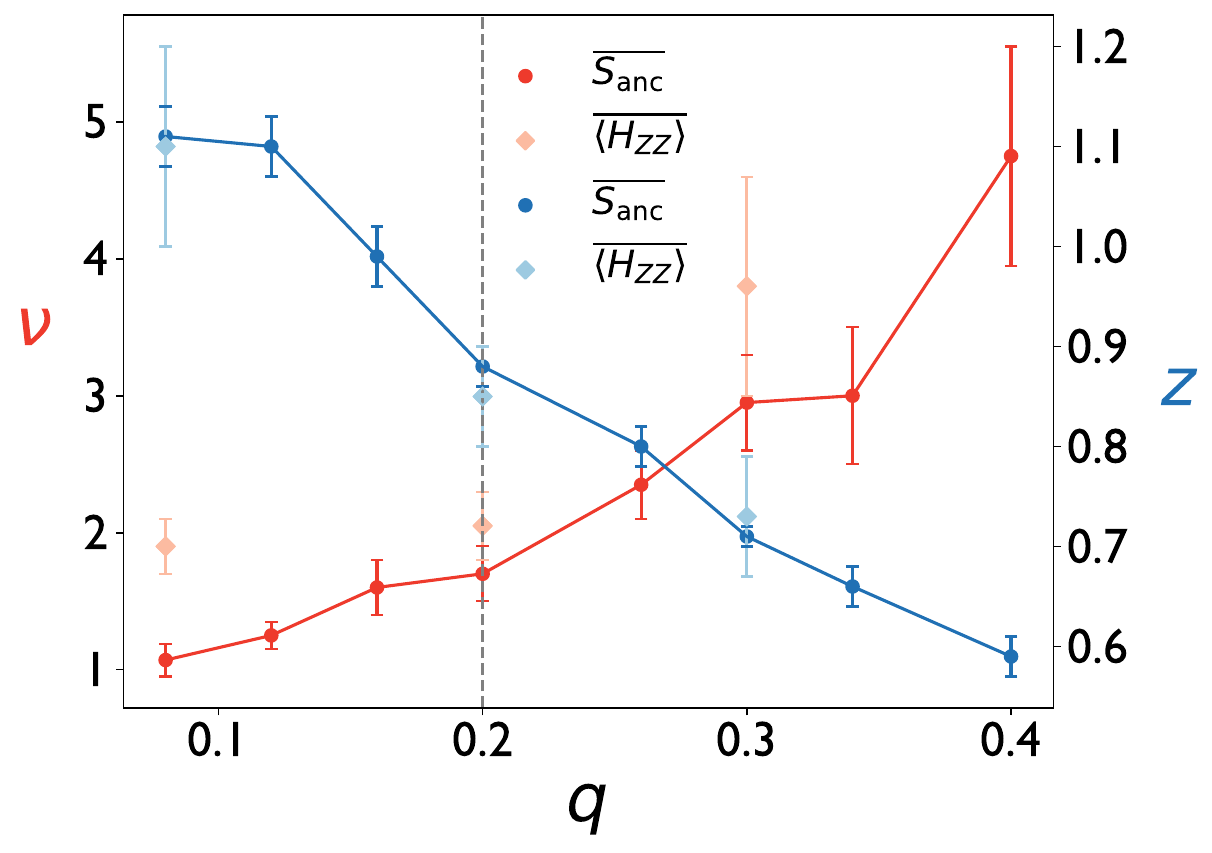}
    \caption{\textbf{Evolution of critical exponents (quantum model).} Correlation length exponents $\nu$ (red) and dynamical exponents $z$ (blue) for the entanglement (darker circles) and control transitions (lighter diamonds) are plotted along with their error bars as a function of the local correction density $q$. The dashed line at $q=0.2$ represents the transition studied in previous sections.}
    \label{fig:nu-z}
\end{figure}

Figs.~\ref{fig:phase-diagram} and \ref{fig:nu-z} largely support this intuition.
In Fig.~\ref{fig:phase-diagram} we plot a portion of the phase diagram as a function of $p$ and $q$, focusing on $\overline{S_{\rm anc}}$ as a reliable probe of the entanglement transition. 
As expected, we see that $p_c$ (red points) drifts to larger values as $q$ decreases, implying that the control circuit must be applied more frequently in order to drive the entanglement transition.
To test whether the locations of the entanglement and control transitions continue to coincide as $q$ is varied, we also plot the values of $p_c$ extracted from $\overline{\braket{H_{ZZ}}}$ (purple points) for $q=0.3$ and $0.08$, in addition to the value obtained for $q=0.2$ (dashed line) in the previous section.
We find that the transitions coincide up to the resolution of our small-size numerics for $q=0.3$ and $0.2$, but that the transitions appear to separate for $q=0.08$.

In Fig.~\ref{fig:nu-z} we plot the critical exponents $\nu$ (red) and $z$ (blue) along the phase boundary as $q$ is varied for both the entanglement (darker circles) and control transitions (lighter diamonds).
For both transitions, we see the expected trend of increasing $\nu$ and decreasing $z$ as $q$ increases.
We also see that the dynamical exponents of the two transitions remain very close to one another, and well within error bars, as $q$ varies.
Notably, for small $q \lesssim 0.16$, the dynamical exponent $z$ becomes greater than $1$, signaling that the transitions are recovering a ``short-range" nature with critical correlations spreading ballistically or slower.
The values of $\nu$ for the entanglement and control transitions coincide up to the resolution of our numerics for $q=0.2$ and $0.3$ where the transitions are ``long-range" with $z<1$, while they differ by several error bars at $q=0.08$ where the transitions have $z\approx 1.1$.
Making a more conclusive statement on the respective values of $p_c$, $z$, and $\nu$ for the two transitions requires system sizes beyond those accessible to our exact numerics.
This may be achievable by applying matrix product state techniques from the controlled side of the transition, which is a worthwhile subject for future work.

\section{Discussion and Outlook}
\label{sec: Conclusion}

We have studied the impact of collective measurements on quantum control and entanglement dynamics in a model inspired by Rydberg atom quantum simulators and by the classical notion of probabilistic control of chaos.
The critical points and critical exponents extracted from the various metrics considered in our study at $q=0.2$ are collected for reference in Table~\ref{tab:exponents}.
With the aid of a numerically tractable classical limit, we identified a control transition driven by collective measurements and local adaptive correction that features a dynamical critical exponent $z<1$, indicating ``superballistic'' spreading of correlations at criticality.
The correlation-length and dynamical exponents of the transition obtained in the classical and quantum limits of the model are in agreement.
In the quantum limit of the model, we also establish the existence of an entanglement transition witnessed by two independent measures, $\overline{\mathcal I_3}$ and $\overline{S_{\rm anc}}$.
The estimates of the entanglement critical point extracted from both measures are indistinguishable from the estimated location of the control transition to within the resolution of our finite-size numerics.
Furthermore, both entanglement metrics estimate a correlation length critical exponent $\nu$ between $1.4$ and $1.9$, which is consistent with the value $2.05(25)$ estimated from the control order parameter.
Finally, the dynamical critical exponent $z<1$ estimated from the dynamics of $S_{\rm anc}$ is consistent with the value estimated from the dynamics of the control order parameter $H_{ZZ}$ in both the quantum and classical limits.

Upon varying $q$, we find that the entanglement and control transitions retain a ``long-range" nature with $z<1$ down to $q\sim 0.16$, below which they revert to a ``short-range" nature with $z\gtrsim 1$.
In the long-range regime with $z<1$, the locations and critical exponents of the control and entanglement transitions coincide up to the resolution of our numerics.
In the short-range regime, we find evidence that the transitions begin to take on distinct locations and critical properties, including different values of the correlation length exponent $\nu$.
We conjecture that the long-range character of the transition for $q\gtrsim 0.16$ is essential to the apparent coincidence of the transitions.
Our model allows access to kinetically constrained MIPTs as $q\to 0$ and infinitely fast transitions with $z\to 0$ as $q \to 1$. 
The ability to explore these regimes as a function of $p$ and $q$ is an exciting outcome of our study.

\begin{table}[t!]
    \centering
    \begin{tabular}{ccccc}
    \hline \hline
         & $p_c$ & $\nu$ & $\beta$ & $z$\\
         \hline
        $\overline{\braket{H_{ZZ}}}$ (classical) & 0.49(1) & 2.3(2) & 0.17(2) & 0.8(1)\\
        $\overline{\braket{H_{ZZ}}}$ (quantum) & 0.46(5) & 2.05(25) & 0.32(4) & 0.85(5) \\
        $\overline{\mathcal I_3}$ & 0.44(2) & 1.6(2) &  & \\
        $\overline{S_{\rm anc}}$ & 0.41(2) & 1.7(2) &  & 0.88(2)\\
        \hline\hline
    \end{tabular}
    \caption{\textbf{Transition points and critical exponents at $\bm{q=0.2}$}. Extracted from the various metrics considered in this paper.
    }
    \label{tab:exponents}
\end{table}

It is interesting to consider the implications of our results for the topology of the control-entanglement phase diagram---in particular, do the control and entanglement transitions coincide, such that the control criticality serves as a reliable witness of the entanglement transition?
Two scenarios present themselves.
First, the control and entanglement transitions indeed coincide, and their critical properties match.
Second, the control and entanglement transitions occur very close to one another but do not coincide.
The first scenario is consistent with the observation that control and entanglement transitions can coincide in the presence of a nonlocal control protocol~\cite{Iadecola2023,SierantTurkeshi2023,Pan24}.
Furthermore, in cases where the two transitions occur nearby but do not coincide, typically it is still possible to deduce that the transitions are different by observing a difference in critical properties.
For example, in Ref.~\cite{ODea22} the entanglement and absorbing-state transitions can be distinguished by their value of $z$ even when the transitions are brought close together.
Typically, in this scenario, the entanglement transition belongs to the universality class of the MIPT~\cite{Zabalo20} with $z=1$ and $\nu\approx 1.3$, while universality of the control transition is often governed by classical physics, e.g.~a random walk in Ref.~\cite{Iadecola2023} and the directed-percolation or parity-conserving universality classes in Refs.~\cite{ODea22,Ravindranath22}, all of which feature $z>1$.
In the present case, the entanglement and control transitions have consistent dynamical critical exponents.
Moreover, the correlation length critical exponents $\nu$ for the two transitions are also not distinguishable within the resolution of our numerics, except at small local correction density $q$.
This further supports the hypothesis that the transitions coincide for sufficiently large $q \gtrsim 0.16$.

If the transitions do not coincide, then the entanglement transition manifests critical exponents associated with a nonstandard MIPT.
Given the dynamical exponent $z<1$, it is natural to look to the MIPT with long-range scrambling gates studied in Ref.~\cite{Block22}.
There, two-qubit Clifford gates are applied between qubits a distance $r$ apart with probability $P(r)\sim 1/r^\alpha$, with $\alpha=0$ corresponding to all-to-all interactions.
Ref.~\cite{Block22} found an MIPT with exponents that vary continuously as a function of $\alpha$, with standard short-range MIPT criticality emerging for $\alpha \gtrsim 3.0$.
Intriguingly, an MIPT with $z\approx 0.88$ (matching the dynamical exponent of the entanglement transition studied here at $q=0.2$) occurs when $\alpha \approx 2.9$.
The corresponding correlation length critical exponent $\nu \approx 1.4$, which is within error bars of the correlation length exponent $\nu=1.6(2)$ determined from the TMI, but just outside error bars of the value $\nu=1.7(2)$ determined from the ancilla entropy.
However, this comparison should be made with caution as MIPTs in Clifford circuits are known to exhibit distinct universality relative to, e.g., Haar random circuits~\cite{Zabalo20}.
Moreover, even if the transitions ultimately do not coincide, the fact that their dynamical exponents are so similar seems to suggest that they are intertwined in some way.

This points to one worthwhile direction for future work, which is to study how the standard MIPT is affected by the inclusion of magnetization measurements.
This would be useful in establishing a baseline to which the critical properties of the entanglement transition studied here can be compared.
Another particularly interesting question is how to understand the emergence of a dynamical exponent $z<1$ in the entanglement transition.
This dynamical exponent indicates that collective measurements are enhancing teleportation events similar to how rare regions can within a local projective measurement context \cite{shkolnik2024infinitelyfastcriticaldynamics}.
For example, is it possible to connect the critical dynamics at the MIPT to a model with explicit long-range interactions?
Further, the question about the nature of the volume-law phase in the presence of magnetization measurements needs further exploration.
For example, in monitored random circuits the entanglement in the volume-law phase acquires a subleading system-size dependent correction related to the roughness exponent of a directed polymer in a random environment~\cite{Li23}.
Does a new exponent emerge in the presence of long-range measurements?

Our results demonstrate how collective measurements in adaptive quantum dynamics can be harnessed to robustly control onto dynamical states, which is relevant for the development of efficient quantum algorithms and error correction protocols in NISQ devices.

\begin{acknowledgments}
We acknowledge helpful discussions with Wen-Wei Ho, Nishad Maskara, Lorenzo Piroli, and Sarang Gopalakrishnan.
We especially thank Sriram Ganeshan and Michael Buchhold for numerous inspiring conversations and collaboration on related topics.
This work was supported in part by the National Science Foundation under grants DMR-2143635 (T.I.) and DMR-2238895 (J.H.W.).
J.H.P.~was supported by the Office of Naval Research grant No.~N00014-23-1-2357.
This work was initiated and performed in part at the Aspen Center for Physics, which is supported by National Science Foundation Grant No.~PHY-1607611.
This work was also supported in part by the International Centre for Theoretical Sciences (ICTS) via participation in the program ``Periodically and quasi-periodically driven complex systems" (code: ICTS/pdcs2023/6).
This research was supported in part by grant NSF PHY-2309135 to the Kavli Institute for Theoretical Physics (J.H.P.).
\end{acknowledgments}

\appendix

\section{Magnetization Measurement}
\label{sec:Magnetization_Measurement}
\begin{figure*}
    \centering
    \includegraphics[width=2\columnwidth]{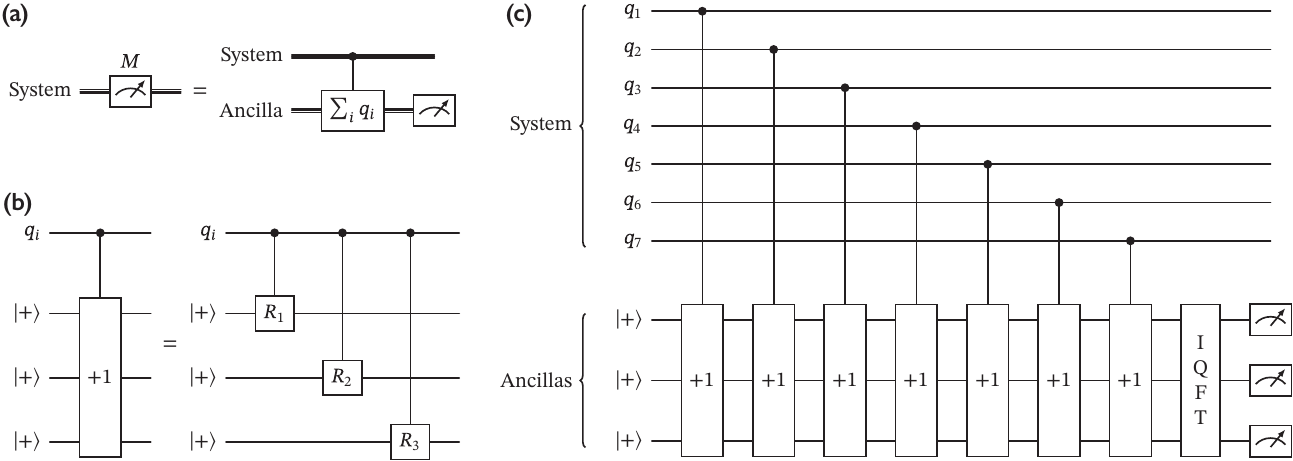}
    \caption{(a) Performing the magnetization measurement, which involves a projector with a large rank can be thought of by controlling an ancillary system by the collective quantity of interest and then projectively measuring the ancillary system. (b) To update ancillary qubits, we conditionally apply $R_j$ to the $j$th ancilla. In the Fourier basis, this updates the ancillary state by $q_i$. (c) The full protocol to measure the magnetization of a system of $L=7$ qubits. It requires $3 = \log_2(L+1)$ ancillas and an inverse Quantum Fourier Transform at the end to allow the measurement to be done in the $Z$-basis of each ancilla individually. }
    \label{fig:magnetization_meas}
\end{figure*}

To measure magnetization in a system of qubits, we are inspired by Refs.~\cite{McArdle19,Botelho22} (see also Refs.~\cite{Wang21,Piroli24}).
The operator we are measuring is
$$
M = \sum_{i=1}^L Z_i.
$$
For ease in this section, we will use the equivalent number operator, where if $Z_i \ket{q_i} = (1 - 2q_i) \ket{q_i}$ for $q_i = 0,1$, then $n_i \ket{q_i} = q_i \ket{q_i}$.
We want to accomplish the procedure of taking a state $\ket{\Psi}$ and projecting it into the number sector $n$ using the projector
$$
P_n = \sum_{\sum_i q_i = n} \ket{q_1q_2\cdots q_L}\bra{q_1 q_2 \cdots q_L}
$$
with probability $\braket{\Psi | P_n | \Psi}$.
\begin{equation} \label{eq:magmeas}
    \ket{\Psi} \mapsto \frac{P_n\ket{\Psi}}{\sqrt{\braket{\Psi|P_n|\Psi}}}, \quad \text{with probability $\braket{\Psi| P_n |\Psi}$.}
\end{equation}

To accomplish this, we break it down into a few steps.

\textbf{Step 1: Couple and measure an ancillary system}. There are $L+1$ number states ($0$ to $L$), and thus, if we can couple in an $d > L$ dimensional qudit, we can projectively measure it to perform the collective measurement (illustrated in Fig.~\ref{fig:magnetization_meas}(a)).  

To define this coupling, we need a unitary gate $U_\mathrm{mag}$ with the following property (where $\ket{\cdot}_a$ represents the ancillary system)
$$
 U_\mathrm{mag} \ket{q_1 \cdots q_L} \otimes \ket{0}_{a} = \ket{q_1 \cdots q_L} \otimes \big\lvert\sum_i q_i\big\rangle_a.
$$
If we measure the ancilla, we use the projector $P^{\mathrm{anc}}_n = \openone \otimes \ket{n}\!\bra{n}_a$, to obtain
\begin{equation}
    P_n^{\mathrm{anc}} U_\mathrm{mag} \ket{\Psi} \otimes \ket{0}_a = P_n \ket{\Psi} \otimes \ket{n}_a,
\end{equation}
with probability $\braket{\Psi| P_n |\Psi}$, achieving Eq.~\eqref{eq:magmeas}.

To accomplish this with quantum hardware, we can choose $d = 2^{\lceil{\log_2(L+1)}\rceil}$, so that we have $\lceil \log_2(L+1)\rceil$ ancillary qubits.

\textbf{Step 2: Designing $U_\mathrm{mag}$.}
There are a number of ways this can be designed, but here we describe one that uses the Fourier basis from the Quantum Fourier Transform \cite{nielsen_quantum_2011}.

Using $N = \lceil \log_2(L+1)\rceil$ qubits, define the Fourier states
\begin{align}
\ket{Q}_a & \equiv \frac1{\sqrt{2^N}} \sum_{a_1,\cdots a_N=0}^1 e^{2\pi i Q a / 2^N} \ket{a_1} \otimes \cdots \otimes \ket{a_N}, \\ & = \frac1{\sqrt{2^N}} \bigotimes_{j=1}^N \left(\ket{0} + e^{2\pi i Q / 2^j}\ket{1} \right).
\end{align}
where $a_j = 0,1$ and $a = a_1 + a_2 2^1 + \cdots a_N 2^{N-1}$ and thus $Q = 0, \ldots, 2^N - 1$.

We therefore initialize the system in $\ket{0}_a = \bigotimes_{j=1}^N \ket{+}$, and we can update the state using controlled phases between system qubit $i$ and ancilla $j$
\begin{equation}
    R_m(i,j) \ket{q_i} \otimes \ket{a_j} = e^{2\pi i q_i a_j / 2^m} \ket{q_i} \otimes \ket{a_j},
\end{equation}
where all products should be understood to be right-to-left.
By performing an $R_j(i,j)$ controlled phase on the $j$th ancilla, one can update the state
$$
\prod_{k=1}^N R_k(i,k) \ket{q_i} \otimes \ket{n}_a = \ket{q_i} \otimes \ket{n + q_i \!\!\! \mod d}_a.
$$
An example of this is in Fig.~\ref{fig:magnetization_meas}(b) labeled as the $+1$ gate (so named because it either leaves the ancilla register alone or adds one).
If we subsequently apply this to all qubits, each gets added so that
$$
\prod_{i=1}^L \prod_{k=1}^N R_k(i,k) \ket{q_1\cdots q_L} \otimes \ket{0}_a = \ket{q_1\cdots q_L} \otimes \big\lvert \sum_i q_i \big\rangle_a.
$$
We have thus created exactly the unitary
$$
U_\mathrm{mag} = \prod_{i=1}^L \prod_{k=1}^N R_k(i,k).
$$
To measure the magnetization then, assuming we can only measure single qubits, we need to convert from the Fourier basis to the qubit basis, which is accomplished with an inverse Quantum Fourier Transform (IQFT) purely on the ancillas.
If we then measure the ancillas, the readout will be the magnetization in binary.
This full protocol for a system of size $L=7$ is illustrated in Fig.~\ref{fig:magnetization_meas}(c).
(Note that similar circuits are presented in Refs.~\cite{Wang21,Piroli24}, and that the approach outlined here is equivalent to quantum phase estimation for the operator $M$ applied to a state that is not an $M$ eigenstate.)

After this is done, the ancilla register can be re-used in either of two ways: Performing a Quantum Fourier transform again and tracking how it is updated by the next $U_\mathrm{mag}$ or resetting all ancillas into the $\ket{+}$ state to restart the procedure.

\section{Control onto General Short Orbits}
\label{sec: General Control}

The control protocol described in Sec.~\ref{sec: Control} readily generalizes to define control onto more general periodic patterns with $r$-site unit cells. 
(The case discussed above is an example in which $r=2$.) 
These patterns generate what Ref.~\cite{Iadecola20} refers to as ``short orbits" of length at most $2^r$ under reversible CA dynamics.
Such orbits generically occur in translation-invariant reversible CA dynamics and can persist under appropriately designed stochastic CA dynamics~\cite{Shiraishi18,Iadecola20}.

The generalization proceeds as follows.
Control can be quantified by a Hamiltonian analogous to Eq.~\eqref{eq:HZZ} with $Z_iZ_{i+2}$ replaced by $Z_iZ_{i+r}$, so that the control protocol must find domain walls on $r$ inequivalent sublattices instead of two.
To determine which of the $2^r$ possible patterns is closest, a measurement of the total magnetization on each of the $r$ sublattices can first be performed, from which the minimal Hamming distance can then be determined.
Domain wall correction then proceeds by measuring $Z_{i}Z_{i+r}$ with probability $q$, and, if a domain wall is present, measuring $Z_i$ and then flipping whichever bit would reduce the Hamming distance to the chosen point on the orbit.

This raises the question of whether it might be possible to control onto longer orbits, e.g. ones whose lengths scale polynomially in $L$ rather than being $O(1)$. However, in this case the problem of determining the ``closest" orbit point is more complicated and naively requires measuring extensively many operators to determine. One possible way to circumvent this is to control onto a single state in the orbit. However, if the control is successful in driving the system to one of these states, a subsequent application of the chaotic CA will generate dynamics onto a different state in the orbit. From the point of view of the control map, this state appears to have extensively many defects. As soon as the control map tries to ``correct" these, it pushes the system off the orbit (assuming that $q\neq 1$). This further underscores the necessity of a multistable control protocol, i.e.~one for which each point on the target orbit is a fixed point.

\section{Finite-Size Decay of Observables}
\label{sec: Sticky Orbits}

In this section, we illustrate the finite-size decay of the control order parameter and half-chain entanglement entropy and discuss how this decay can be removed.

\begin{figure}[tb!]
    \centering
    \includegraphics[width=\columnwidth]{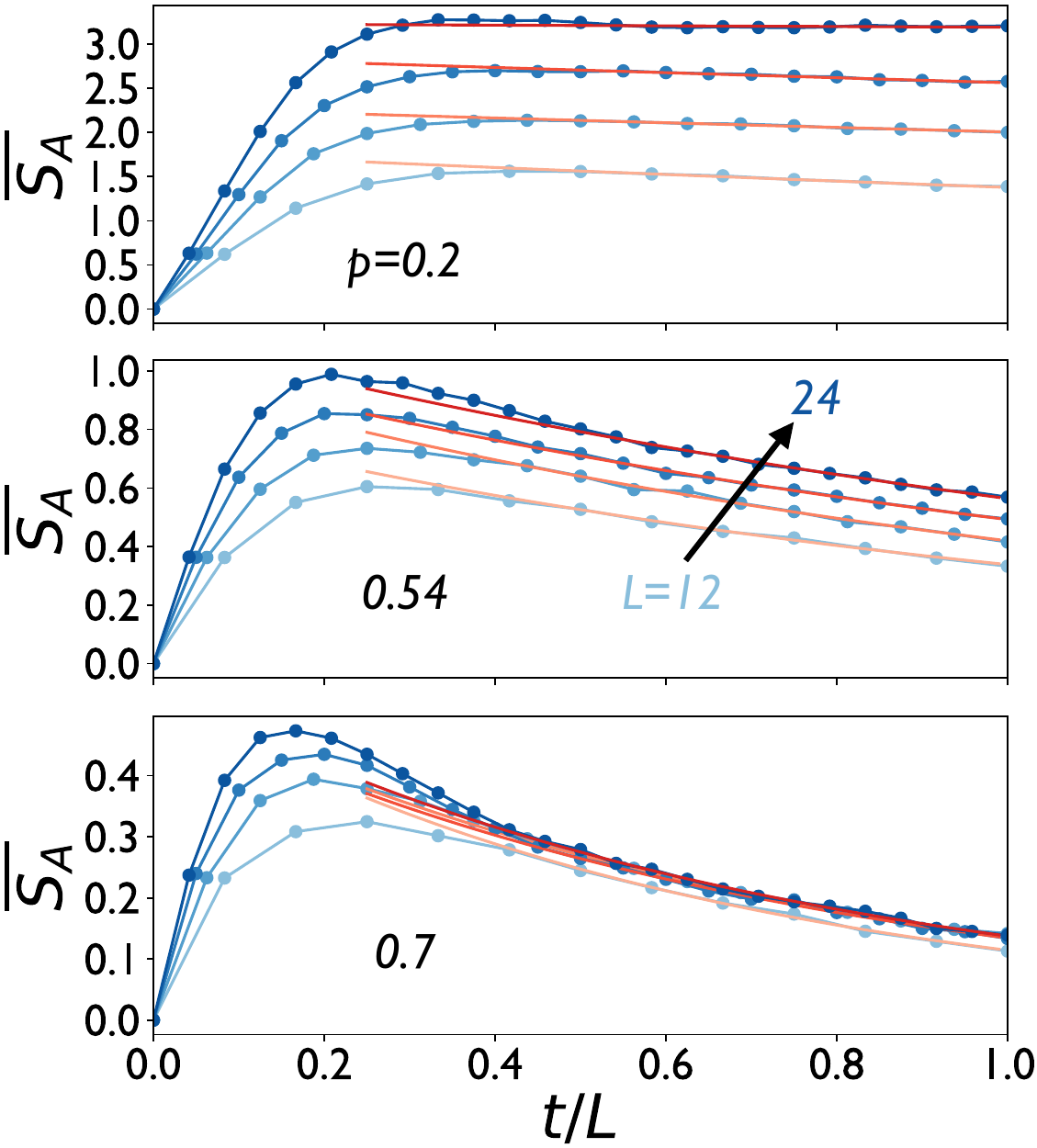}
    \caption{\textbf{Decay of the von-Neumann entanglement entropy.} Dynamics of the von-Neumann entanglement entropy for several system sizes at three values of $p$ representative of the volume-law phase (top), the critical regime (middle), and the area-law phase (bottom). Red lines indicate best fits to an exponential decay of the form \eqref{eq:SA-decay} for times $t\geq L/2$.}
    \label{fig:half-cut-evol}
\end{figure}

The entanglement transition studied in the main text manifests similar finite-size effects as the order parameter discussed in Sec.~\ref{sec: Control Transition} and offers a good example of how to extract the steady-state thermodynamic limit.
First, this decaying behavior in Fig.~\ref{fig:half-cut-evol} is expected since the system becomes fully disentangled when it becomes ``accidentally trapped" on the target orbit due to a rare circuit realization.
Rather than approaching a $p$- and $L$-dependent saturation value $S_{\infty}(p,L)$, as it would in an MIPT for example, the late-time entanglement entropy takes the schematic form in Eq.~\eqref{eq:SA-decay} in which $S_{\infty}(p,L)$ is modulated by an exponentially decaying envelope characterized by a decay rate $\Gamma(p,L)$.

Entanglement dynamics of the form \eqref{eq:SA-decay} is evident in Fig.~\ref{fig:half-cut-evol}, which plots the dynamics of $\overline{S_A}$ at three values of $p$ representative of the volume-law phase (top), the area-law phase (bottom), and the regime near the critical point (middle).
For all three $p$ values, the late-time entanglement entropy for $t\sim L$ decays with time rather than showing a tendency to saturation. (We have verified this statement by carrying out simulations to much longer times $\sim 10L$, where the system still exhibits clear exponential decay in $t/L$ and no saturation is observed.)
The red lines show fits to the schematic exponential decay form of Eq.~\eqref{eq:SA-decay}, focusing on times $t\geq L/2$.
The decay parameter $\Gamma(p,L)$ appears to decrease with $L$ in the volume law phase, as can be seen in the top panel of Fig.~\ref{fig:half-cut-evol}.
In contrast, near the critical point and deep in the area-law phase, $\Gamma(p,L)$ appears to be independent of $L$, as can be seen in the bottom two panels of Fig.~\ref{fig:half-cut-evol}.

This late-time behavior of the entanglement still allows for a suitable definition of different dynamical entanglement phases based on the system-size scaling of $\overline{S_A}$ at any fixed sufficiently late time $t\sim L$.
In our analysis of the entanglement entropy scaling, we evaluate $\overline{S_A}$ at time $t=L$ and see clear evidence of an entanglement transition between a volume-law phase where $\overline{S_A}$ grows with $L$ and an area-law phase where it is $L$-independent.
Varying the precise time $t$ at which the entanglement entropy is evaluated does not have a strong effect on the system-size scaling, provided it is sufficiently late that the system has reached the exponential decay regime visible in Fig.~\ref{fig:half-cut-evol}.

\begin{figure}[b!]
    \centering
    \includegraphics[width=\columnwidth]{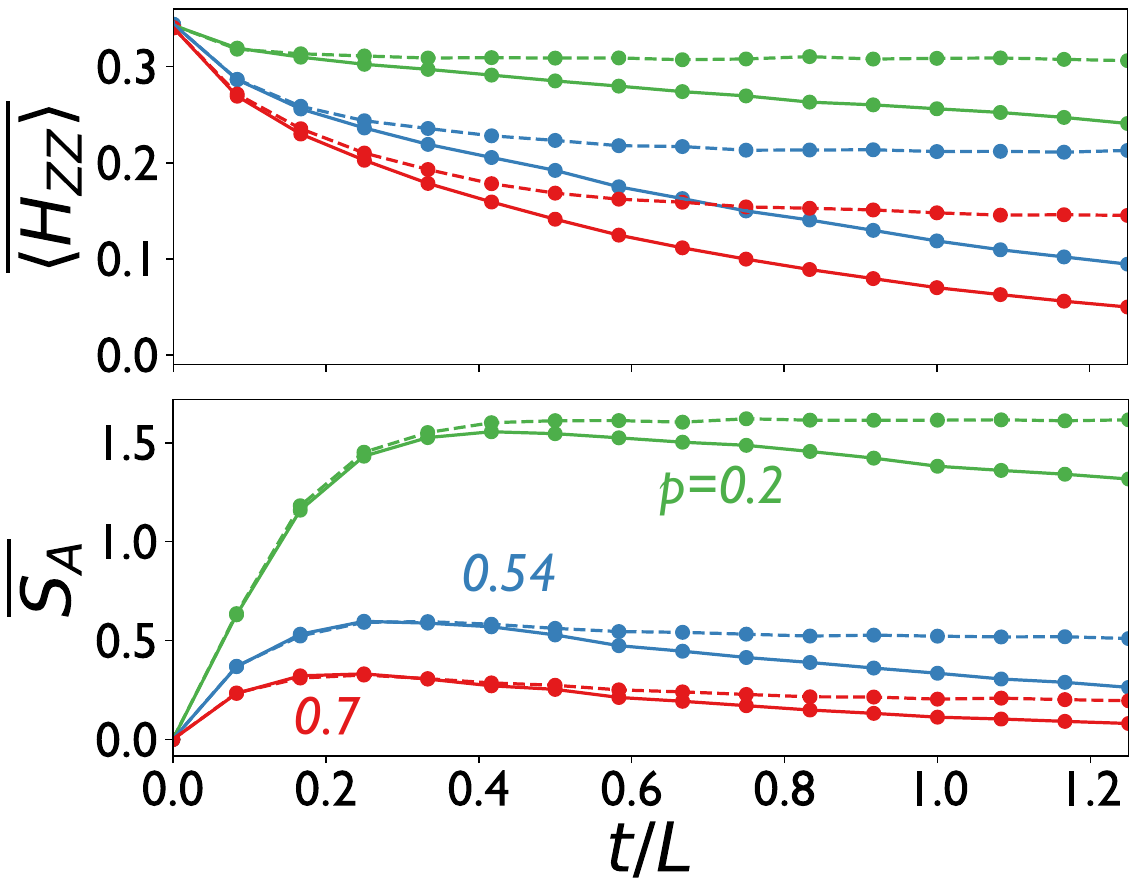}
    \caption{
    \textbf{Dynamics of perturbed and unperturbed model}. Order parameter (top) and half-chain entanglement entropy (bottom) at $L=12$ and $q=0.2$ for several values of $p$.
    Solid lines represent results obtained using the dynamical protocol studied in the main text, while dashed lines represent results obtained by modifying this protocol to remove the target orbit from the chaotic circuit.
    }
    \label{fig:orbit-appendix}
\end{figure}

However, this decay regime can be avoided by adding a sub-extensive perturbation to the dynamics.
In Fig.~\ref{fig:orbit-appendix}, we plot the unperturbed quantities at $L=12$ for three values of $p$ (solid lines).
As in the main text, data points represent averages over $10^4$ samples, and error bars (smaller than the points) indicate the standard error of the mean over these samples.
Both the control order parameter and the entanglement entropy continue to decay beyond the time $t=L$ at which these quantities were extracted to perform the finite-size scaling analysis in the main text.

To remove this decay, we add one more step to the chaotic circuit of Sec.~\ref{sec: Circuit}.
After applying $U_{\rm PXP}$ and $U_{\sigma}(\theta)$, we apply a (conditional) bit flip gate of the form \eqref{eq:correct} on a single qubit, chosen at random each time the chaotic circuit is applied;
The rationale for this is that a finite fraction of the wave function gets trapped on the vacuum orbit and if we introduce a perturbation that is sub-extensive and kicks us off of the orbit, the system will not fully decay in the putative volume-law phase (at small system sizes).
The result is illustrated with dashed lines in Fig.~\ref{fig:orbit-appendix}, which display clear saturation as expected in a standard dynamical phase.
While we have not verified that the transition persists in the presence of this modification, a similar strategy was employed for $\beta$-adic R\'enyi circuits in Refs.~\cite{Iadecola2023,LeMaire24,Allocca24,Pan24}, where a clear second-order phase transition is present.
It is an interesting question for future work to determine the amount by which the system can be ``pushed" off of the target orbit while preserving the transition.

\bibliography{refs, zotero_refs}




\end{document}